\documentclass[referee]{raa}           
\usepackage{graphicx,times}
\usepackage{natbib}
\usepackage{amssymb,amsmath}
\bibpunct{(}{)}{;}{a}{}{,}
\usepackage{CJK}
\usepackage[graphicx]{realboxes}

\usepackage[a4paper=true,pagebackref=true]{hyperref}
\hypersetup{pdftitle = The title of my PDF, pdfauthor = My name, pdfsubject= The subject, pdfkeywords = keyword1 keyword2 keyword3} 
\hypersetup{colorlinks = true, linkcolor = green, anchorcolor = red, citecolor = blue, filecolor = red, pagecolor = red, urlcolor = red}

\newcommand{\NHn}{NH$_3$}
\newcommand{\NnHn}{N$_2$H$^+$}

\newcommand{\CHnCN}{CH$_3$CN}

\newcommand{\CnHn}{C$_{3}$H$_{2}$}

\newcommand{\HnO}{H$_2$O}

\newcommand{\HnCO}{H$_2$CO}

\newcommand{\Msun}{M$_\odot$}

\newcommand{\kms}{{km~s$^{-1}$}}
\newcommand{\unitMsunyr}{M$_\odot$yr$^{-1}$}
\newcommand{\HCOn}{{HCO$^{+}$}}
\newcommand{\HnCOn}{{H$^{13}$CO$^{+}$}}

\begin{document}

   \title{Studying Infall in Infrared Dark Clouds with Multiple {\HCOn} Transitions
%$^*$
%\footnotetext{\small $*$ Supported by the National Natural Science Foundation of China.}
}

 \volnopage{ {\bf 20XX} Vol.\ {\bf X} No. {\bf XX}, 000--000}
   \setcounter{page}{1}

   \author{{\begin{CJK*}{UTF8}{gkai}XIE Jinjin（谢津津）\end{CJK*}}\inst{1,2,3}, Jingwen Wu
      \inst{2,1}, Gary~A. Fuller\inst{3,4}, Nicolas Peretto\inst{6}, Zhiyuan Ren\inst{1},  Longfei Chen
      \inst{1}, {\begin{CJK*}{UTF8}{gkai}Yaoting Yan (闫耀庭)\end{CJK*}}\inst{7}, Guodong Li\inst{1,2}, Yan Duan\inst{1,2}, Jifeng Xia\inst{1,2}, {\begin{CJK*}{UTF8}{bkai}Yongxiong Wang（王永雄）\end{CJK*}}\inst{3}, {\begin{CJK*}{UTF8}{bkai}Di Li（李菂）\end{CJK*}}\inst{1,2,5}
   }

   \institute{ National Astronomical Observatories, Chinese Academy of Sciences, Beijing 100101, China; {\it jingwen@nao.cas.cn; dili@nao.cas.cn}\\
        \and
             University of Chinese Academy of Sciences, Beijing 100049, China\\
	\and
	  Jodrell Bank Centre for Astrophysics, Department of Physics \& Astronomy, The University of Manchester, Manchester M13 9PL, United Kingdom\\
	  \and
I. Physikalisches Institut, University of Cologne, Z\"ulpicher Str. 77, 50937 K\"oln, Germany\\
\and 
NAOC-UKZN Computational Astrophysics Centre, University of KwaZulu-Natal, Durban 4000, South Africa\\
\and
School of Physics \& Astronomy, Cardiff University, Queen\textquotesingle s Building, The Parade, Cardiff, CF24 3AA, the United Kingdom\\
\and
Max-Planck-Institut f\"{u}r Radioastronomie, Auf dem H\"{u}gel 69, 53121 Bonn, Germany\\
\vs \no
   {\small Received 20XX Month Day; accepted 20XX Month Day}
}

\abstract{We investigate the infall properties in a sample of 11 infrared dark clouds (IRDCs) showing blue-asymmetry signatures in {\HCOn}$J\,$=1--0 line profiles. We used JCMT to conduct mapping observations in {\HCOn}$J\,$=4--3 as well as single-point observations in {\HCOn}$J\,$=3--2, towards 23 clumps in these IRDCs. We applied the HILL model to fit these observations and derived infall velocities in the range of 0.5$-$2.7\,{\kms}, with a median value of 1.0\,{\kms}, and obtained mass accretion rates of 0.5-14\,$\times 10^{-3}$\,{\unitMsunyr}. These values are comparable to those found in massive star forming clumps in later evolutionary stages. These IRDC clumps are more likely to form star clusters. {\HCOn}$J\,$=3--2 and {\HCOn}$J\,$=1--0 were shown to trace infall signatures well in these IRDCs with comparable inferred  properties. {\HCOn}$J\,$=4--3, on the other hand, exhibits infall signatures only in a few very massive clumps, due to smaller opacties. No obvious correlation for these clumps was found between infall velocity and the {\NHn}$/$CCS ratio.
}

   \authorrunning{XIE J. J. et al. }            %author_head in even pages
   \titlerunning{Infall in IRDCs with {\HCOn}}  % title_head in odd pages
   \maketitle

%________________________________________________ sections below
% 
\section{Introduction}           %% first-level sections will be auto-capitalized
\label{sect:intro}

Mass accumulation by inward gravitational motions is a basic step in models of star formation \citep[e.g.][] {1969larson,1987anglada,2001bonnell,2018motte}. Infall motions often are revealed in observations with moderate optical depth molecular lines, which show blue-shifted self-absorption dips at frequencies where optically thin lines peak, produced by the temperature gradients of dense cores and the infalling gas \citep[e.g.][] {1986walker,2003wu}. Multiple transitions of many molecules such as CS, CO, {\HnCO}, {\HCOn}, HCN, {\CHnCN}, and {\CnHn} have been investigated to search for infall signatures in various environments in low mass star forming regions \citep[e.g.][] {1995myers,1997mardones,1998tafalla,2016keown}, as well as massive star forming regions \citep[e.g.][] {1998zhang,2003wu,2005fuller,2008velusamy,2010barnes,2016qin}. 

The comparisons of different tracers including multiple transitions of the same tracer have been made through both observations and simulations, to explore which tracers are more efficient to reveal infall signatures in what kind of sources. Higher-$J$ transitions, such like $J\,$=3--2 and $J\,$=4--3, of HCN and {\HCOn} are considered as good tracers for dense clusters based on their infall asymmetries in lines generated in numerical simulations \citep{2014chira}. Observationally, HCN $J\,$=3--2 has been considered as a very good infall tracer in a dense clump sample associated with {\HnO} masers \citep{2003wu}. The    $J\,$=1--0 transition of HCN was found to have the strongest infall signatures \citep{2010stahleryen} in simulations although the line shapes depend 
on the viewing angle \citep{2012smith}. Compared to HCN, {\HCOn} changes less drastically in abundances in chemical models at temperatures ranging from 10\,K to 40\,K \citep{2012vasyunina}, which is the temperature range of dark clouds. Thus, {\HCOn} is potentially better for comparing infall in different star forming environments as it is less affected by chemistry. Among {\HCOn}, the lowest transition, {\HCOn} $J\,$=1--0, showed the most blue-asymmetric profiles towards a sample of 77 candidate high mass protostellar objects (HMPOs) \citep{2005fuller}, which is consistent with the results of numerical simulations that {\HCOn} $J\,$=1--0 is a better indicator of collapse in high mass star formation \citep{2013smith} where the gas is not dense enough to excite the higher-$J$ transitions.

Being cold and dense, infrared Dark Clouds (IRDCs) are believed to represent the initial conditions of massive star formation and the formation of the associated stellar clusters \citep[e.g.][] {2000carey,2006rathborne,2009perettofuller,2018motte}. Some pilot {\HCOn} observations towards two massive IRDCs have revealed that the clumps/cores in there are undergoing rapid collapse \citep{2013peretto,2018contreras}. Considering IRDCs are at an earlier stage than HMPOs or UCHII massive clumps, it raises the question of what molecular or ion line tracers can best trace infall in IRDCs? Specifically, will lower-$J$ transitions with smaller critical density (like {\HCOn}$J\,$=1--0) reveal infall signature better, or higher-$J$ transitions are needed to probe collapse in these IRDCs? In addition, it is important to understand the properties of infall in the IRDC phase compared with infall in more evolved regions.
%? Is it different from infall happening in later stages of massive star formation?

Recently, in a {\HCOn}$J\,$=1--0 mapping survey towards a sample of 27 IRDCs (Peretto et al. in preparation), a significant fraction presented infall signatures in their line profiles. In this paper, we report our follow-up study on a sub-sample of these IRDCs, all of which show infall signatures in {\HCOn}$J\,$=1--0 in at least one clump in the IRDCs. We mapped 11 IRDCs with {\HCOn}$J\,$=4--3, and made single-point observations with {\HCOn}$J\,$=3--2, towards 23 {\HCOn}$J\,$=4--3 peak positions, using James Clerk Maxwell Telescope (JCMT). The aims of the study are to use multiple {\HCOn} transitions to fit and constrain models of infall motion, to understand infall properties in IRDCs, and to investigate if higher-$J$ transitions of {\HCOn} can also probe infall and reveal infall properties well in these {\HCOn}$J\,$=1--0 selected IRDCs . 

The organization of the rest of the paper are the follow: We describe source selection and observations in sect. \ref{sect:Obs}. The observational results and and data analysis are presented in sect. \ref{sect:results}. We discuss infall tracers and infall properties of IRDCs in sect. \ref{sect:discussion}, and summarise our conclusions in sect. \ref{sect:concl}.

% Authors can give a citation as `\citealt{Michel+etal+1992}'.
% You may also use \cite, \citep and \citet for citation, and use Table~1
% or Figure~1 and so forth. Using \ref and \label for cross-references of
% Tables/Figures is a good way in adjusting/adding/removing text, tables or
% figures.

%\begin{figure} 
 %  \centering
  % \includegraphics[width=14.0cm, angle=0]{sample2_fig1.eps}
  % \begin{minipage}[]{85mm}
   %\caption{The YFOSC spectra of six new quasars. From the left to right, the red dashed lines mark the wavelengths of Ly$\alpha$, SiIV and CIV emission lines at the estimated redshift for five $z>3.6$ quasars, while for SDSS J113816.85+045023.6 they mark the wavelengths of CIV and CIII]. } 
%\end{minipage}
   %\label{Fig1}
   %\end{figure}

\section{Target Selection and Spectroscopic Observations}
\label{sect:Obs}

The sources in this work are a sub-sample of an IRAM 30\,m {\HCOn} $J\,$=1--0 survey towards 27 IRDCs (Peretto et al. in preparation). The parent sample was selected to span a range of geometries and a factor of $\sim$100 in mass range from $\sim$200 to $\sim$2$\times$10$^{4}${\Msun}, to cover a representative sample of IRDCs. From the parent sample we selected 11 IRDCs who show blue asymmetry indicative of infall in {\HCOn} $J\,$=1--0 in at least one clump. We made follow-up higher-$J$ {\HCOn} observations and we fit multiple transitions with models, in order to study infall properties in these candidate infalling IRDCs. The basic information of the sources are listed in Table~\ref{tab:sources}. 

All the sources have been mapped in {\HCOn} $J\,$=1--0 with IRAM 30\,m telescope (half power beam width (HPBW): 29"; $\eta_{mb}$=0.75). Since {\NnHn} $J\,$=1--0 has been found in good agreement with {\HnCOn} $J\,$=1--0 in both velocity and line width \citep{2005fuller},  we use {\NnHn} $J\,$=1--0, which was observed simultaneously with {\HCOn} $J\,$=1--0, to locate the central velocity of the cloud and to identify infall signature for all the  {\HCOn} transitions. The IRAM 30\,m observations and data reduction of {\HCOn} $J\,$=1--0 and {\NnHn}$J\,$=1--0 of the 27 IRDCs will be presented in a separated paper (Peretto et al. in preparation).

Eleven IRDCs were mapped in {\HCOn}$J\,$=4--3 with Heterodyne Array Receiver Programme (HARP) \footnote{https://www.eaobservatory.org/jcmt/instrumentation/heterodyne/harp/.} \citep{harp} (HPBW:: 14"; $\eta_{mb}$=0.63) on the 15\,m James Clerk Maxwell Telescope (JCMT)\footnote{The JCMT is operated by the EAO on behalf of NAOJ; ASIAA; KASI; CAMS as well as the National Key R\&D Program of China (No. 2017YFA0402700). Additional funding support is provided by the STFC and participating universities in the UK and Canada.} on Mauna Kea, Hawaii during August and September 2016 (Project M16BP081), and August 2017 (Project M17BP087). The Auto-Correlation Spectral Imaging System (ACSIS) spectrometer was used. Maps of 180" by 180" (except one source SDC18.624, whose map size is 240"$\times$180") were made with RASTER scans with T$_{sys}$ ranging from 364\,K to 613\,K (with an average of 443\,K), under weather Band 3 and Band 4 with 0.08 $< \tau_{225\,GHz} < $ 0.20. 

Single-point observations were carried out in {\HCOn}$J\,$=3--2 towards the {\HCOn}$J\,$=4--3 peak positions with JCMT RxA3m frontend and ACSIS backend \footnote{https://www.eaobservatory.org/jcmt/instrumentation/heterodyne/rxa/} (HPBW: 20"; $\eta_{mb}$=0.57) from April to June 2018 under Project M18AP073. No {\HnCOn}$J\,$=3--2 observations were available due to a low LO current at 260\,GHz. The molecule line frequencies are listed in Table~\ref{tab:molecular_line}. Each position was observed in position-switching mode (GRID) for an integration time of 300\,s in weather Band 4 (0.12 $< \tau_{225\,GHz} < $ 0.2). An off-position (+600", +600") (J2000) from the observed position was also observed. The system temperatures ranged from 778\,K to 1356\,K, with an average of 988\,K. For both the HARP and RxA3m observations, ACSIS was configured to cover 250\,MHz wide windows, 8192 channels in each window, resulting to a velocity resolution of $\sim$ 0.02\,{\kms} for HARP and $\sim$ 0.03\,{\kms} for RxA3m. All data the HARP and RxA3 observations were smoothed with a Gaussian kernel which has the FWHM equals 5.9 channels, resulting a velocity resolution of 0.15\,{\kms} and 0.20\,{\kms} for HARP and RxA3, respectively. The telescope pointing was checked before observing a new source and was checked every 1-1.5 hours, by observing one or more calibration sources in CO(2--1) at 234.591\,GHz and CO(3--2) at 350.862\,GHz for RxA3m and HARP, respectively. The uncertainty in the flux calibration is estimated to be about 10\%.

The HARP and RxA3m data reduction were undertaken using the {\sc Starlink} \citep{starlink} software package {\sc SMURF}, {\sc KAPPA}, and {\sc GAIA}. Each integration was first visually checked. The data were converted to spectral cubes and baselines were subtracted before been written out as FITS format files using standard {\sc Starlink} routines \citep{pipeline}.

All data were converted from the antenna temperature scale T$_{A}^{*}$ to main-beam brightness temperature T$_{mb}$ using T$_{mb}$ = T$_{A}^{*}$/$\eta_{mb}$, where main beam efficiencies are listed in Table~\ref{tab:molecular_line}, as well as the noise levels and the velocity resolutions.

\begin{table}[h!]
\bc
\begin{minipage}[]{100mm}
\caption[]{Physical Parameters of the Observed IRDCs.\label{tab:sources}}\end{minipage}
\setlength{\tabcolsep}{2.5pt}
\small
 \begin{tabular}{lcccc}
  \hline\noalign{\smallskip}
 & R.A.(J2000) & Decl.(J2000) & Distance$^{a}$ &  V$_{LSR}$ \\
Source Name & (hh:mm:ss) & ($^\circ$ ' ")  & (kpc)  & ({\kms}) \\
  \hline\noalign{\smallskip}
SDC18.624-0.070 & 18:25:10.0 & -12:43:45  & 3.50  & 45.6\\
SDC18.888-0.476 & 18:27:09.7 & -12:41:32 & 4.38 & 66.3 \\
SDC22.373+0.446 & 18:30:24.5 & -09:10:34 & 3.61 & 53.0\\
SDC23.367-0.288 & 18:34:53.8 & -08:38:00   & 4.60 & 78.3\\
SDC24.489-0.689 & 18:38:25.7 & -07:49:36  & 3.28  & 48.1\\
SDC24.618-0.323 & 18:37:22.4 & -07:32:18  & 3.04  & 43.4\\
SDC25.166-0.306 & 18:38:13.0 & -07:03:00  & 3.95 & 63.6\\
SDC28.333+0.063 & 18:42:54.1 & -04:02:30   & 4.56  & 79.3\\
SDC35.429+0.138 & 18:55:30.4 & +02:17:10 & 4.67  & 77.0\\
SDC35.527-0.269 & 18:57:08.6 & +02:09:08 & 2.95  & 45.4\\
SDC35.745+0.147 & 18:56:02.6 & +02:34:14  & 5.11 & 83.4\\
\hline
  \noalign{\smallskip}\hline
\end{tabular}
\tablecomments{0.86\textwidth}{The coordinates and V$_{LSR}$ are taken from Peretto et al. (in preparation). $^{a}$The distances are determined from the Galactic Ring Survey (GRS) \citep{2008jackson}. }
\ec
%% place \tablecomments and \tablerefs below \end{center| and \end{center}:
%% you may leave the table-width parameter to editors or set to your actual size
%\tablecomments{0.86\textwidth}{The SDSS $ugriz$ magnitudes are given in AB system 
%and the UKIDSS $YJHK$ magnitudes are given in Vega system.}
\end{table}

\begin{table}
\bc
\begin{minipage}[]{100mm}
\caption[]{Molecular Line Properties.\label{tab:molecular_line}}\end{minipage}
\setlength{\tabcolsep}{2.5pt}
\small
 \begin{tabular}{lc c c c c c c  }
  \hline\noalign{\smallskip}
  Molecular Line &  Frequency (GHz)  & Telescope & $\eta_{mb}$ & HPBW & $\delta$v$^{a}$  ({\kms}) & average rms$^{a}$ (K) & Observation Date \\
  \hline\noalign{\smallskip}
 {\HCOn}$J\,$=1--0 & 89.18852470    & IRAM 30\,m & 0.75 & 29" & 0.16 & 0.07 & 2013$^{b}$ \\
{\HCOn}$J\,$=3--2 & 267.55762590  & JCMT & 0.57 & 20" & 0.20 & 0.33 & 2018 \\
{\HCOn}$J\,$=4--3 & 356.73422300 & JCMT & 0.63 & 14" & 0.15 &0.25  & 2016, 2017\\
\hline
  \noalign{\smallskip}\hline
\end{tabular}
\tablecomments{1.00\textwidth}{$^{a}$ Both $\delta$v and rms are on T$_{mb}$ scale. All data were smoothed with a Gaussian kernel with FWHM equals 5.9 channels. $^{b}$ The IRAM 30\,m observations of {\HCOn}$J\,$=1--0 were made by Peretto et al. (in prep). }
\ec
\end{table}

\section{Results}\label{sect:results}

\subsection{Infall Signature Statistics}

\begin{figure}[h!]
  \includegraphics[angle=0, width=0.98\textwidth]{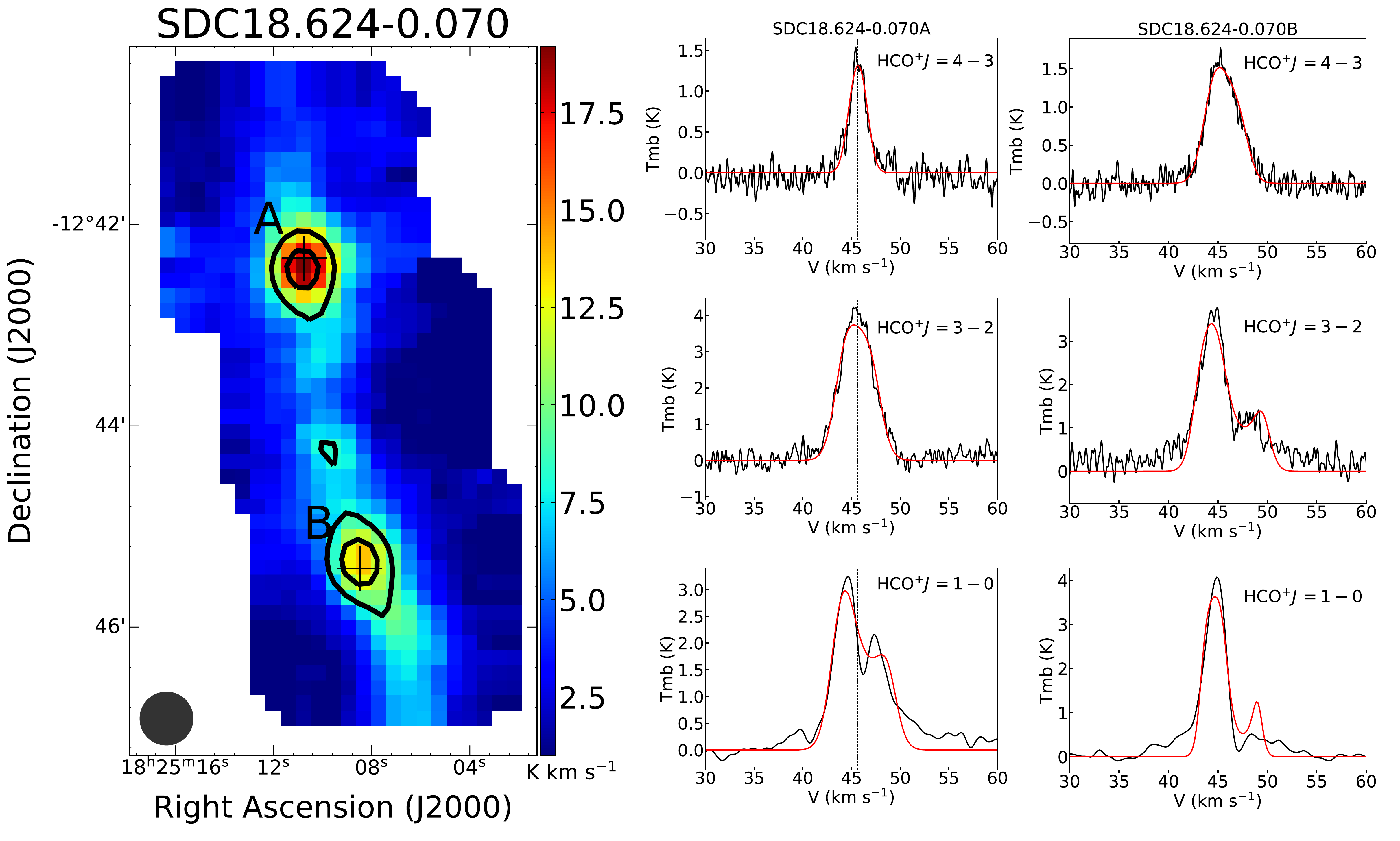}  
    \caption{The integrated intensity map of {\HCOn}$J\,$=1--0 overlaid with {\HCOn}$J\,$=4--3 contours at 0.5, 0.8, and 1.0 of the maximum of the integrated intensity in K {\kms} for SDC18.624. The maximum of the integrated intensity has been stated in Table~\ref{tab:size_luminosity} as the biggest value of $\int T_{MB}dv$ for the source. A and B are the observed positions of {\HCOn}$J\,$=3--2. The spectra are fitted with HILL model. The black emission profiles are from the data, while the red lines represent the fitted profiles. The grey dot at the corner of the emission map represents the IRAM beam size.}
    \label{fig:sdc18p624}
\end{figure}

\begin{figure}
  \includegraphics[angle=0, width=0.98\textwidth]{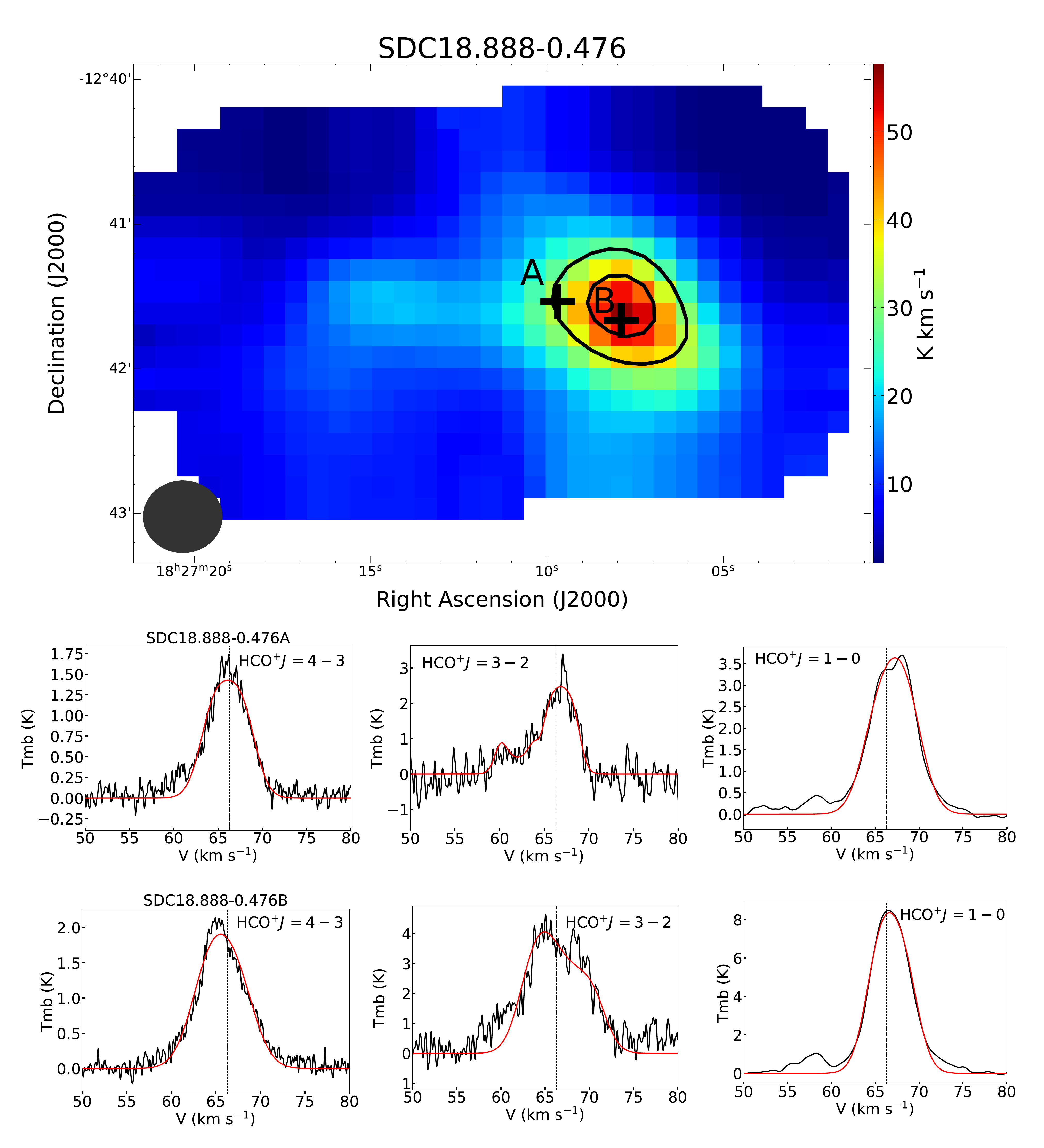}  
    \caption{The integrated intensity map of {\HCOn}$J\,$=1--0 overlaid with {\HCOn}$J\,$=4--3 contours at 0.5, 0.8, and 1.0 of the maximum of the integrated intensity in K {\kms} for SDC18.888. The maximum of the integrated intensity has been stated in Table~\ref{tab:size_luminosity} as the biggest value of $\int T_{MB}dv$ for the source. A and B are the observed positions of {\HCOn}$J\,$=3--2. The grey dot at the corner of the emission map represents the IRAM beam size.}
    \label{fig:sdc18p888}
\end{figure}

We have mapped {\HCOn}$J\,$=4--3 towards 11 IRDCs, which are presented as a contour map overlaid on the emission image of {\HCOn}$J\,$=1--0 in Figure~\ref{fig:sdc18p624} through Figure~\ref{fig:sdc35p745} in \ref{sect:level1}. The {\HCOn}$J\,$=4--3 data have been convolved with the IRAM 30\,m beam size. We identified 23 dense clumps based on {\HCOn}$J\,$=4--3 emission peaks in the maps. These are labelled A through D in the figures. The coordinates of these peaks are tabulated in Table~\ref{tab:infall_velocity}.

The position of the {\HCOn}$J\,$=4--3 peaks generally agree well with {\HCOn}$J\,$=1--0 peaks. An exception is SDC28.333+0.063, in which {\HCOn}$J\,$=4--3 peaks at two positions (B and C in the figure) without corresponding {\HCOn}$J\,$=1--0 peaks. These two clumps may be very small in size and  smoothed out by the much larger beam size in IRAM 30m telescope, or these regions may have quite different physical conditions leading to unusual {\HCOn} $J\,$=4--3 to {\HCOn}$J\,$=1--0 ratio.

We present the spectra of all three transitions towards the peak positions in \ref{sect:level1}. As seen from these spectra, infall signatures can be recognized in most of these clumps, and vary among different transitions. We categorised the profile signatures into five types that suggest different kinematic statuses: (1) double peaks with blue strong, an example can be seen in {\HCOn}$J\,$=3--2 in SDC18.624-0.070B (see Figure~\ref{fig:sdc18p624}); (2) a blue profile with a shoulder, e.g. the profile of {\HCOn}$J\,$=4--3 in SDC18.624-0.070B (see Figure~\ref{fig:sdc18p624}); (3) symmetric profile, e.g. {\HCOn}$J\,$=3--2 in SDC18.624-0.070A (see Figure~\ref{fig:sdc18p624}); (4) a red profile with a shoulder, e.g. {\HCOn}$J\,$=1--0 in SDC18.888-0.476A (see Figure~\ref{fig:sdc18p888}); (5) double peaks with red strong, e.g. {\HCOn}$J\,$=3--2 in SDC18.888-0.476A (see Figure~\ref{fig:sdc18p888}). We summarize the the line profiles of different {\HCOn} transitions for the 23 IRDC clumps in Table~\ref{tab:number_profile}. 

A double peak with a blue peak stronger profile is the best infall signature that indicates infall is ongoing and can be fitted to derive the infall properties. A blue peak with a red shoulder is also an infall signature, although less dramatic, it can still be used to derive infall parameters through modelling. Both profiles are called blue profiles. It is clear both from spectra or from Table~\ref{tab:number_profile} that {\HCOn}$J\,$=1--0 has the best performance in revealing infall in these IRDCs, with 12 out of 23 showing a double peak profile with blue peak strong, and 5 with a blue peak with a shoulder. This is not surprising given this sample has been selected bias to showing the blue asymmetry in {\HCOn}$J\,$=1--0. {\HCOn}$J\,$=3--2 presents a less prominent, yet still comparable, ability to reveal blue profiles, 11 out of 23 having double peaks with the blue peak stronger, and 3 blue peak with a red shoulder profiles. {\HCOn}$J\,$=4--3 only presents profiles with double peaks and the blue peak stronger in 3 clumps, and shows a blue peak with a shoulder in 13 clumps. It is less sensitive in revealing infall in this IRDC sample. More discussion to compare the ability of the 3 transitions to reveal infall will be given in sect.\ref{sect:discussion}.

\begin{table}[h!]
\bc
\begin{minipage}[]{100mm}
\caption[]{The Numbers of Clumps Profile Signatures.\label{tab:number_profile}}\end{minipage}
\setlength{\tabcolsep}{2.5pt}
\small
 \begin{tabular}{lc c c  }
  \hline\noalign{\smallskip}
  Profile Signatures &  {\HCOn}$J\,$=1--0   & {\HCOn}$J\,$=3--2 & {\HCOn}$J\,$=4--3  \\
  \hline\noalign{\smallskip}
Double Peaks with Blue Strong & 12  & 11 & 3  \\
Blue Profile with Shoulder & 5 & 3 & 13\\
Symmetric & 2 & 4 & 2\\
Red Profile with Shoulder & 3 & 1 & 1\\
Double Peaks with Red Strong  & 0 & 4 & 2\\
\hline
  \noalign{\smallskip}\hline
\end{tabular}
\ec
\end{table}

\subsection{Infall Velocity Calculations}
To derive the infall parameters, we adopted the HILL model \citep{2005hill} to fit the spectral lines. 
The HILL model is based on the precursor `two-layer' model \citep{1996myers}, which has been used in many infall studies, though `two-layer' model has been found to underestimate infall velocity by a factor of $\sim$2 \citep{2005hill}. The HILL model assumes a hill shape excitation temperature profile across the cloud, where the centre has the peak excitation temperature $T_{P}$ \citep{2005hill}. The brightness temperature of the molecular line is:
\begin{equation}\small
\begin{aligned}
\Delta T_{B}\left(v\right)=\left(J\left(T_{P}\right)-J\left(T_{0}\right)\right)\left[\left(1-e^{-\tau_{f}\left(v\right)}\right) \Big/ \tau_{f}\left(v\right)-e^{-\tau_{f}\left(v\right)}\left(1-e^{-\tau_{r}\left(v\right)}\right)\Big/\tau_{r}\left(v\right)\right] \\
+ \left(J\left(T_{0}\right)-J\left(T_{b}\right)\right)\left[1-e^{-\tau_{r}\left(v\right)-\tau_{f}\left(v\right)}\right],\label{eq:HILL}
\end{aligned}
\end{equation}
where $T_{B}$ is the brightness temperature defined as $T_{B} = \left(c^{2}\Big/2{\nu}^{2}k\right)I_{v}$, and  $I_{\nu}$ is the specific intensity. The excitation temperature at the outer edges of the core is $T_{0}$. The optical depth $\tau_{f}$ is the optical depth where has the excitation temperature rising along the line of sight and $\tau_{r}$ is the opposite, with a velocity dispersion of $\sigma$ for the entire cloud. We developed a Python fitting code which consists five parameters, i.e. HILL5 model, based on equation Eq.~\ref{eq:HILL} \footnote{This equation is typeset incorrectly in the Astrophysical Journal in \citep{2005hill}. The correct version can be found in the arXiv version in 2004arXiv: astro-ph/0410748.} using Python package {\sc LMFIT}\footnote{https://lmfit-py.readthedocs.io/}, which also provides error estimates. HILL5 is the HILL model with five free parameters, which are optical depth $\tau$, infall velocity $V_{\rm in}$, systematic velocity $V_{\rm LSR}$, velocity dispersion $\sigma$, and excitation temperature $T_{\rm ex}$. HILL5 is considered as the most robust model among the HILL models \citep{2005hill}, and has been applied to calculate infall velocities in various objects such as IRDCs \citep[e.g.][]{2018contreras}, low mass cores \citep[e.g.][]{2017maureira}, and massive starless clumps \citep[e.g.][]{2018calahan}. 

\begin{figure}
  \includegraphics[angle=0, width=0.98\textwidth]{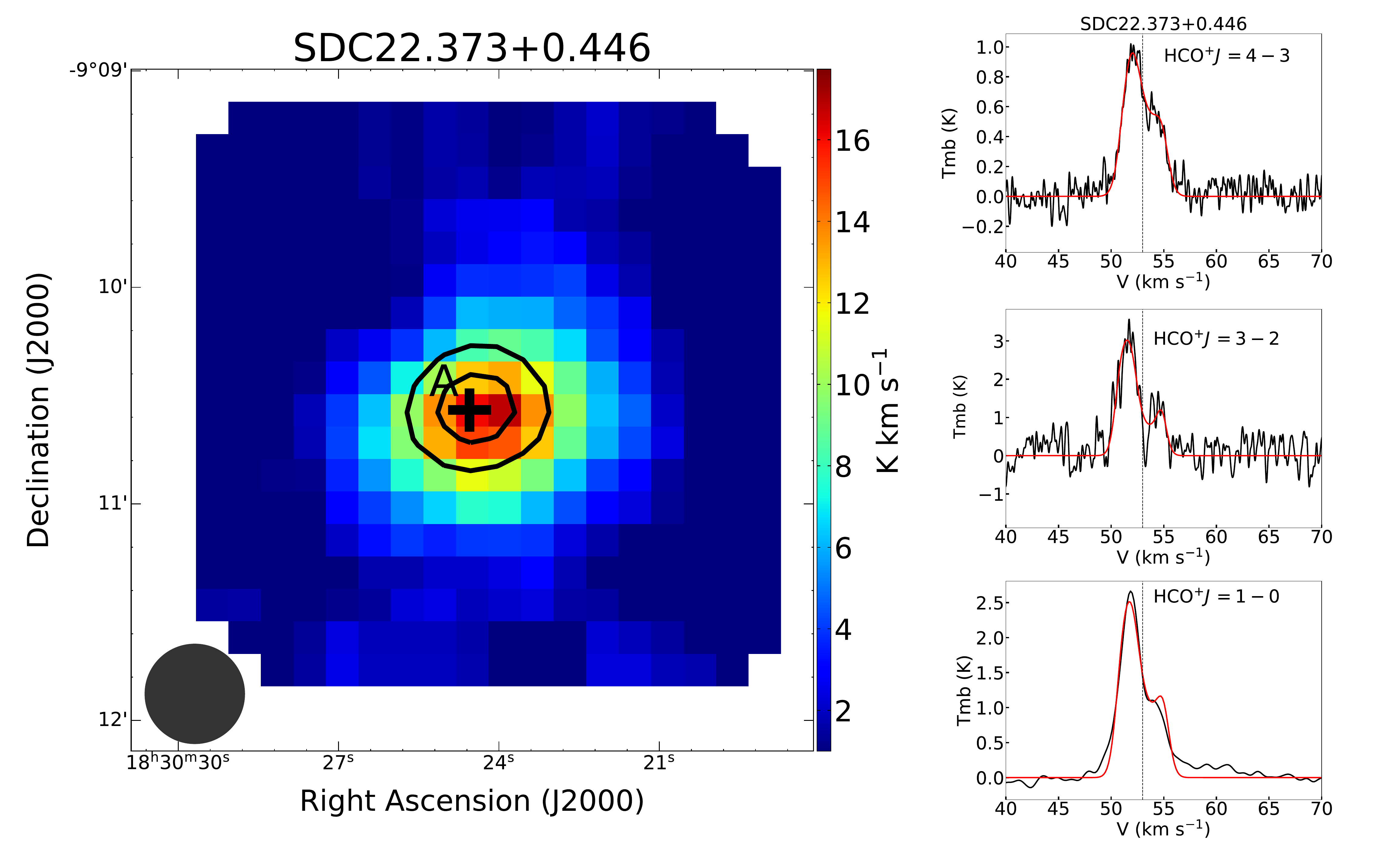} 
    \caption{The integrated intensity map of {\HCOn}$J\,$=1--0 overlaid with {\HCOn}$J\,$=4--3 contours at 0.5, 0.8, and 1.0 of the maximum of the integrated intensity in K {\kms} for SDC22.373. The maximum of the integrated intensity has been stated in Table~\ref{tab:size_luminosity} as the biggest value of $\int T_{MB}dv$ for the source. A and B are the observed positions of {\HCOn}$J\,$=3--2. The grey dot at the corner of the emission map represents the IRAM beam size.}
    \label{fig:sdc22p373}
\end{figure}

\begin{figure}
  \includegraphics[angle=0, width=0.98\textwidth]{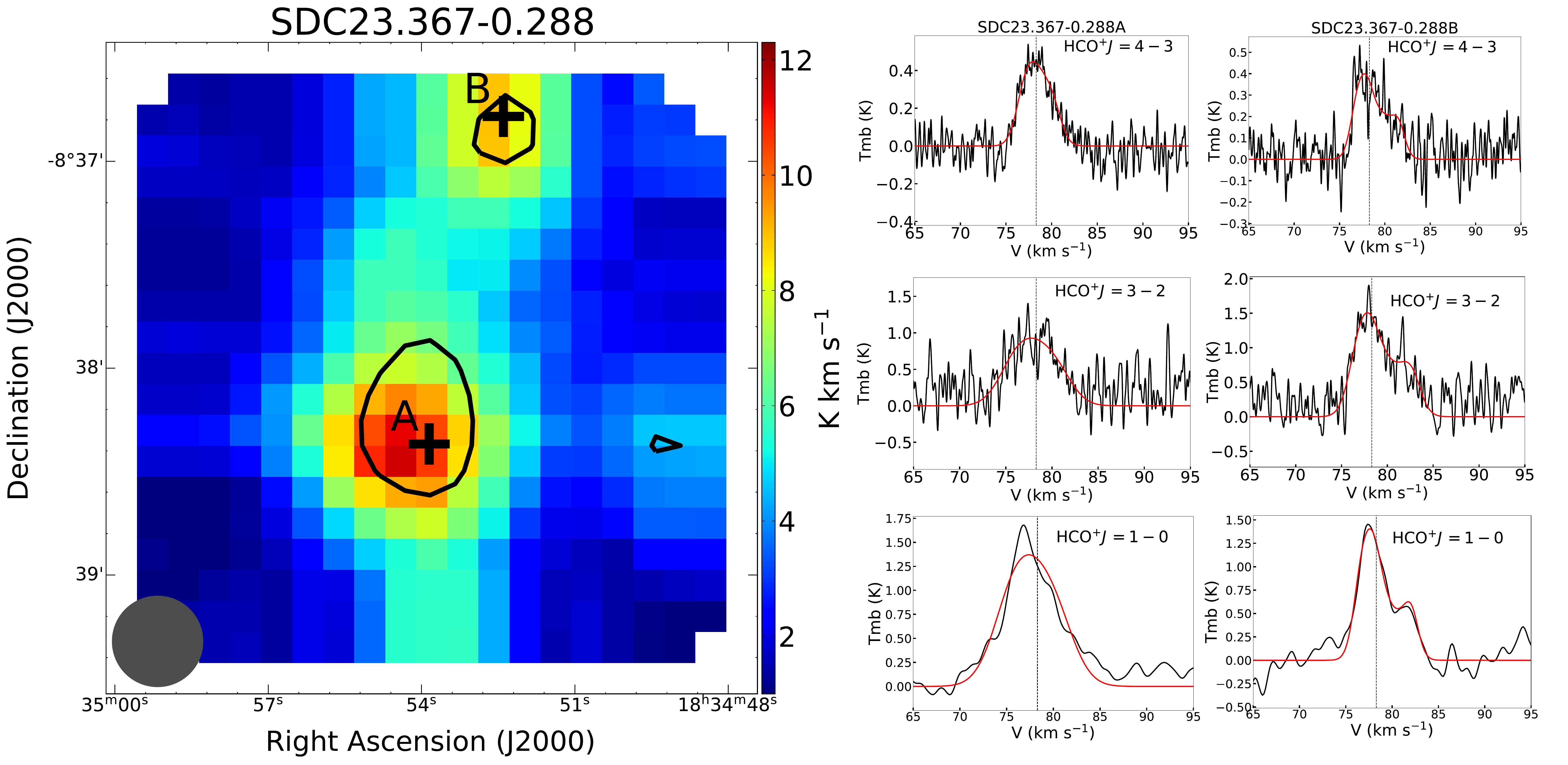} 
    \caption{The integrated intensity map of {\HCOn}$J\,$=1--0 overlaid with {\HCOn}$J\,$=4--3 contours at 0.5, 0.8, and 1.0 of the maximum of the integrated intensity in K {\kms} for SDC23.367. The maximum of the integrated intensity has been stated in Table~\ref{tab:size_luminosity} as the biggest value of $\int T_{MB}dv$ for the source. A and B are the observed positions of {\HCOn}$J\,$=3--2. The grey dot at the corner of the emission map represents the IRAM beam size.}
    \label{fig:sdc23p367}
\end{figure}

\begin{figure}
  \includegraphics[angle=0, width=0.98\textwidth]{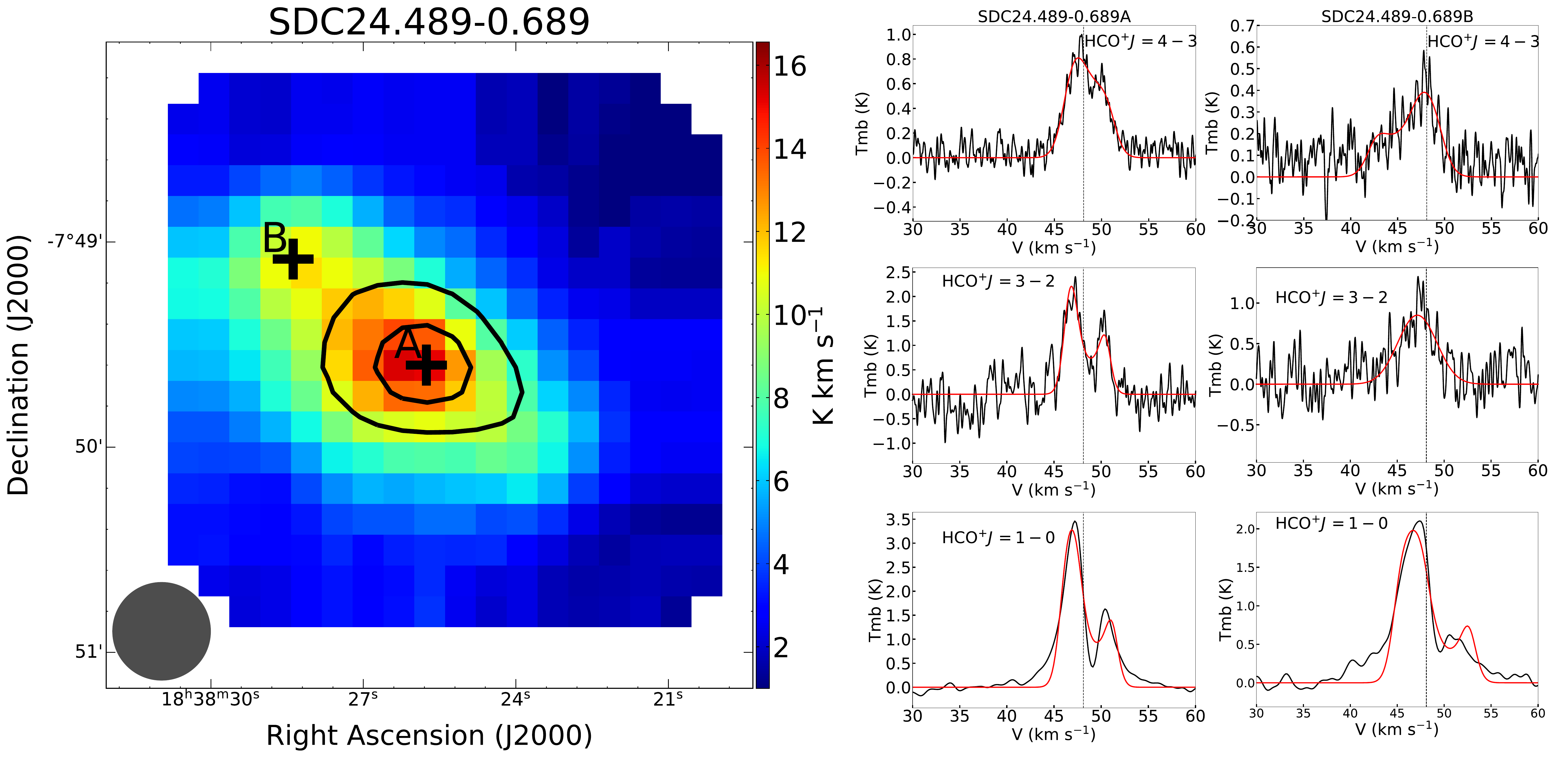}
    \caption{The integrated intensity map of {\HCOn}$J\,$=1--0 overlaid with {\HCOn}$J\,$=4--3 contours at 0.5, 0.8, and 1.0 of the maximum of the integrated intensity in K {\kms} for SDC24.489. The maximum of the integrated intensity has been stated in Table~\ref{tab:size_luminosity} as the biggest value of $\int T_{MB}dv$ for the source. A and B are the observed positions of {\HCOn}$J\,$=3--2. The grey dot at the corner of the emission map represents the IRAM beam size.}
    \label{fig:sdc24p489}
\end{figure}

\begin{figure}
  \includegraphics[angle=0, width=0.98\textwidth]{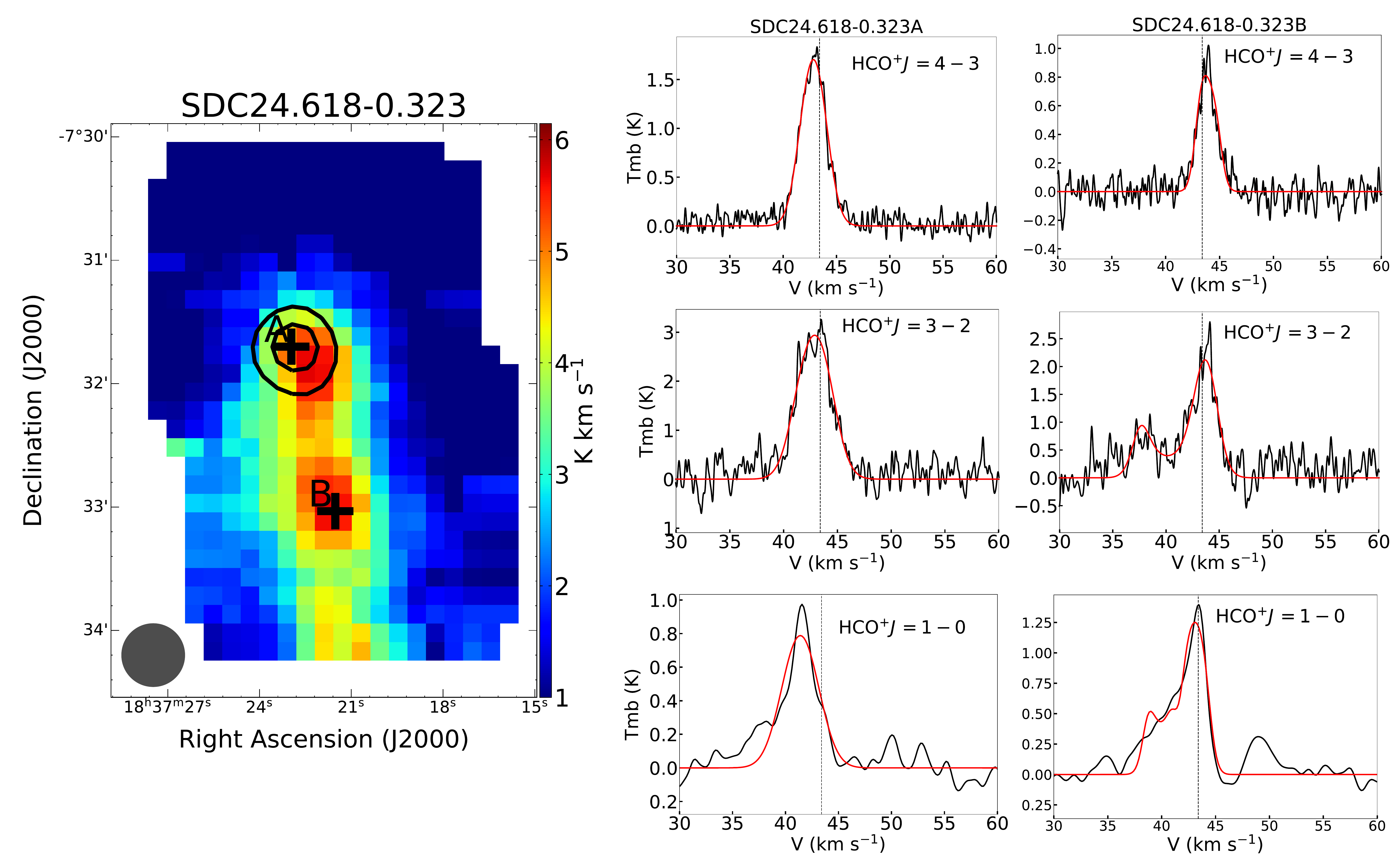} 
    \caption{The integrated intensity map of {\HCOn}$J\,$=1--0 overlaid with {\HCOn}$J\,$=4--3 contours at 0.5, 0.8, and 1.0 of the maximum of the integrated intensity in K {\kms} for SDC24.618. The maximum of the integrated intensity has been stated in Table~\ref{tab:size_luminosity} as the biggest value of $\int T_{MB}dv$ for the source. A and B are the observed positions of {\HCOn}$J\,$=3--2. The grey dot at the corner of the emission map represents the IRAM beam size.}
    \label{fig:sdc24p618}
\end{figure}

\begin{figure}
  \includegraphics[angle=0, width=0.98\textwidth]{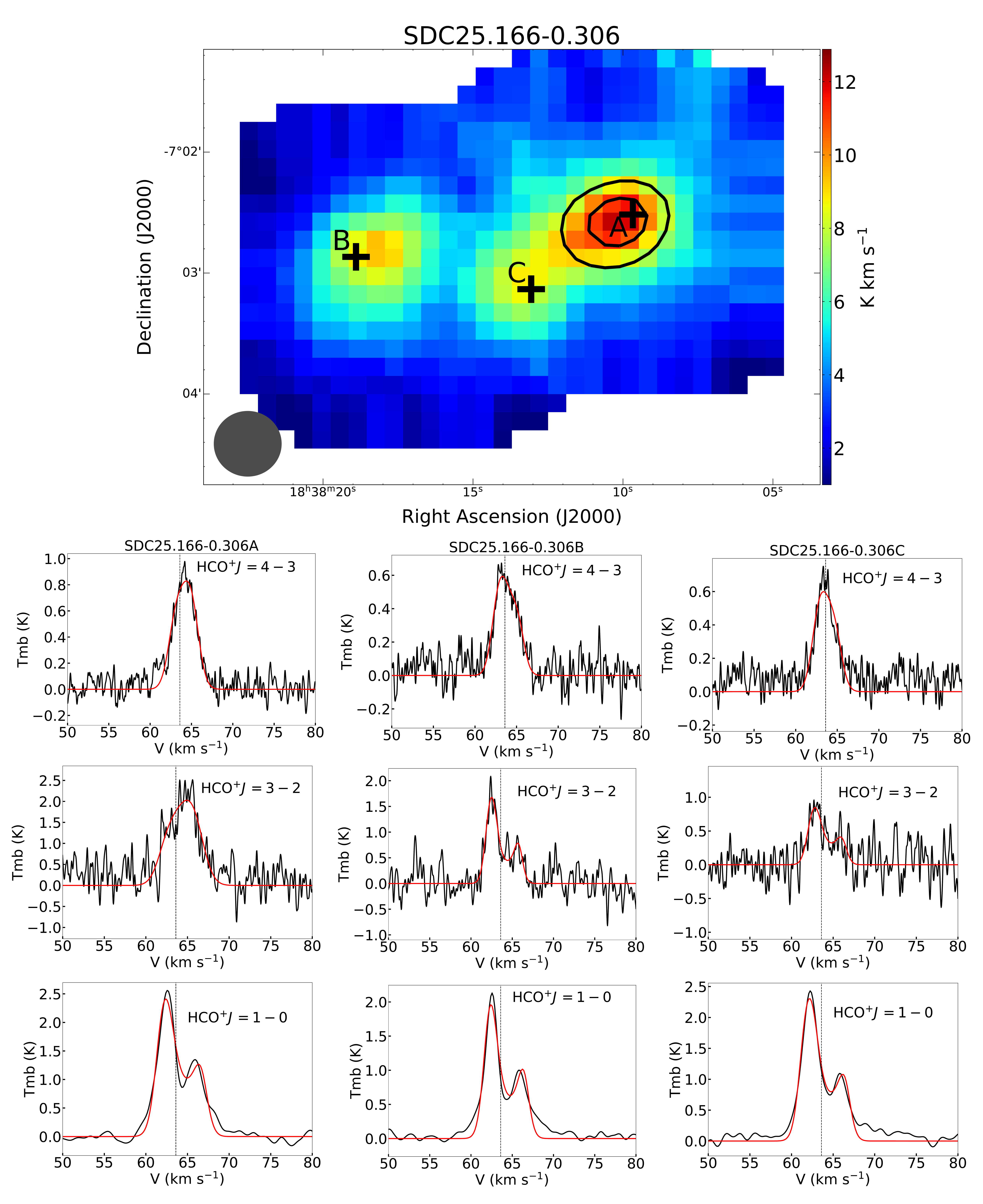} 
    \caption{The integrated intensity map of {\HCOn}$J\,$=1--0 overlaid with {\HCOn}$J\,$=4--3 contours at 0.5, 0.8, and 1.0 of the maximum of the integrated intensity in K {\kms} for SDC25.166. The maximum of the integrated intensity has been stated in Table~\ref{tab:size_luminosity} as the biggest value of $\int T_{MB}dv$ for the source. A, B, and C are the observed positions of {\HCOn}$J\,$=3--2. The grey dot at the corner of the emission map represents the IRAM beam size.}
    \label{fig:sdc25p166}
\end{figure}

\begin{figure}
 \includegraphics[angle=0, width=0.98\textwidth]{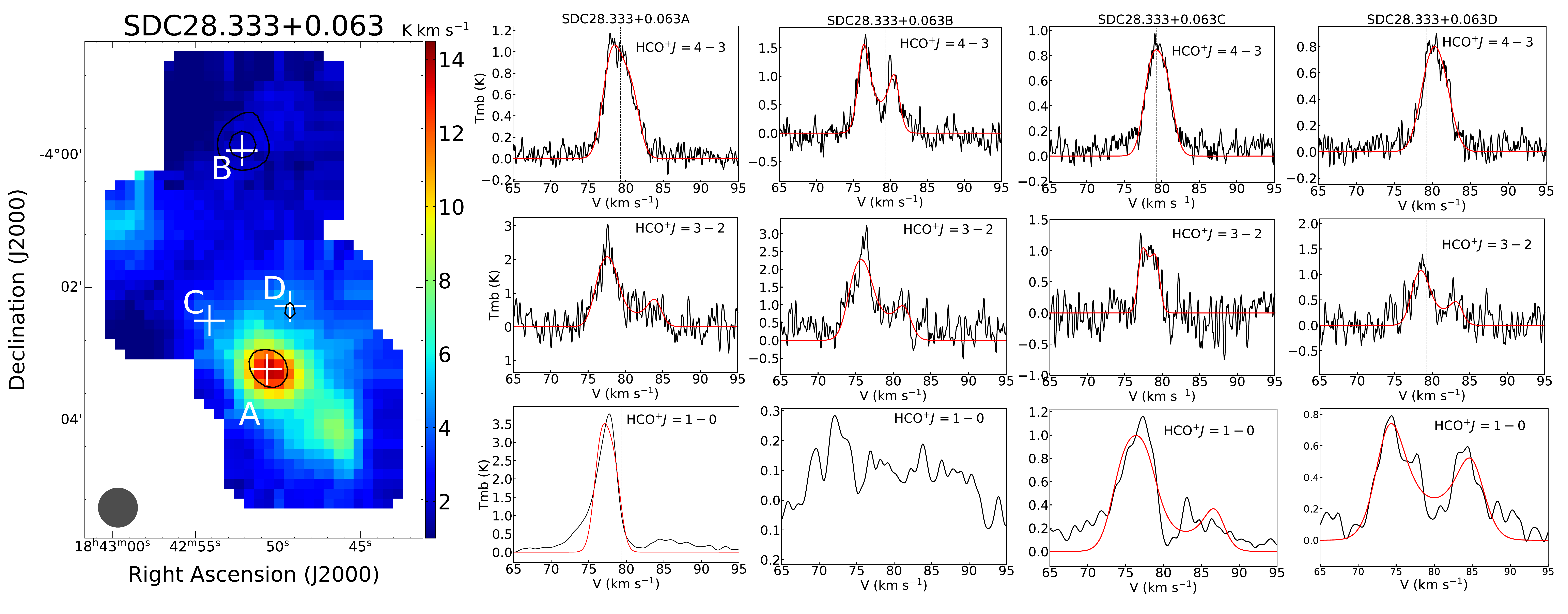} 
    \caption{The integrated intensity map of {\HCOn}$J\,$=1--0 overlaid with {\HCOn}$J\,$=4--3 contours at 0.5, 0.8, and 1.0 of the maximum of the integrated intensity in K {\kms} for SDC28.333. The maximum of the integrated intensity has been stated in Table~\ref{tab:size_luminosity} as the biggest value of $\int T_{MB}dv$ for the source. A, B, C, and D are the observed positions of {\HCOn}$J\,$=3--2. Note that at B position, {\HCOn}$J\,$=1--0 shows no strong emission, despite B position is the emission peak for {\HCOn}$J\,$=4--3 and {\HCOn}$J\,$=3--2 also shows strong emission. The reasons are unknown yet. The grey dot at the corner of the emission map represents the IRAM beam size.}
    \label{fig:sdc28p333}
\end{figure}

\begin{figure}
 \includegraphics[angle=0, width=0.98\textwidth]{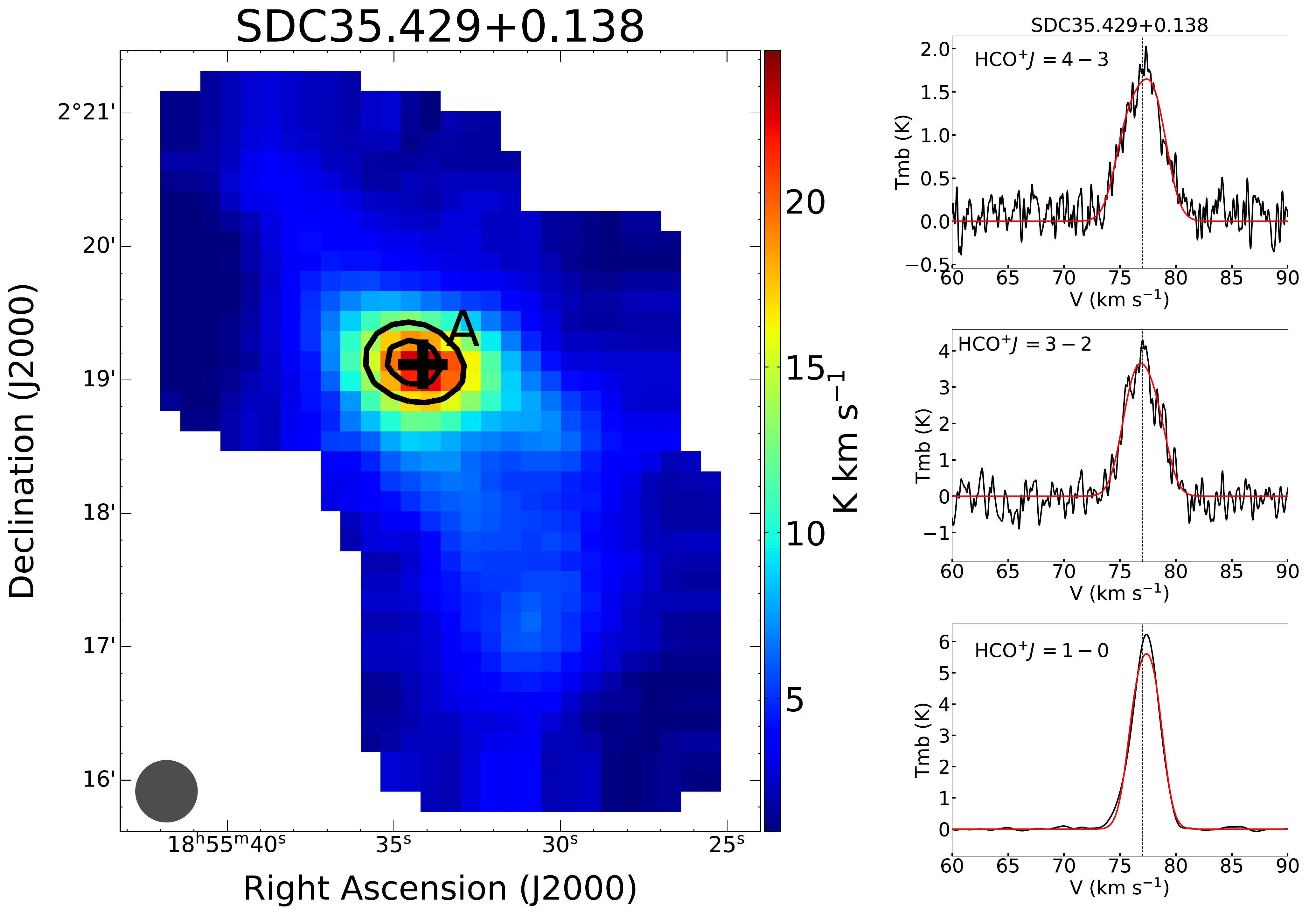} 
    \caption{The integrated intensity map of {\HCOn}$J\,$=1--0 overlaid with {\HCOn}$J\,$=4--3 contours at 0.5, 0.8, and 1.0 of the maximum of the integrated intensity in K {\kms} for SDC35.429. The maximum of the integrated intensity has been stated in Table~\ref{tab:size_luminosity} as the biggest value of $\int T_{MB}dv$ for the source. The grey dot at the corner of the emission map represents the IRAM beam size.}
    \label{fig:sdc35p429}
\end{figure}

\begin{figure}
 \includegraphics[angle=0, width=0.98\textwidth]{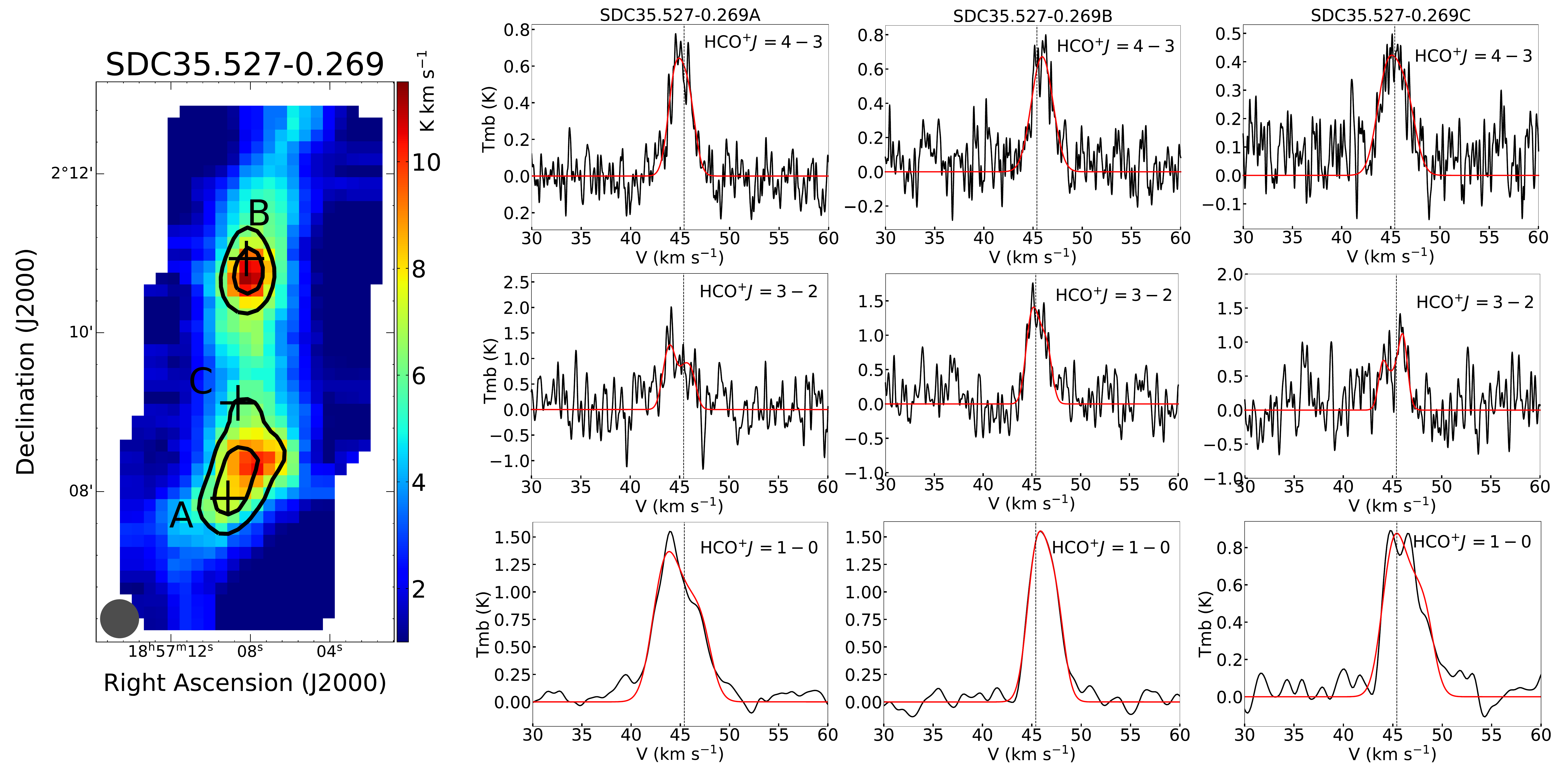} 
    \caption{The integrated intensity map of {\HCOn}$J\,$=1--0 overlaid with {\HCOn}$J\,$=4--3 contours at 0.5, 0.8, and 1.0 of the maximum of the integrated intensity in K {\kms} for SDC35.527. The maximum of the integrated intensity has been stated in Table~\ref{tab:size_luminosity} as the biggest value of $\int T_{MB}dv$ for the source. A, B, and C are the observed positions of {\HCOn}$J\,$=3--2. The grey dot at the corner of the emission map represents the IRAM beam size.}
    \label{fig:sdc35p527}
\end{figure}

\begin{figure}
 \includegraphics[angle=0, width=0.98\textwidth]{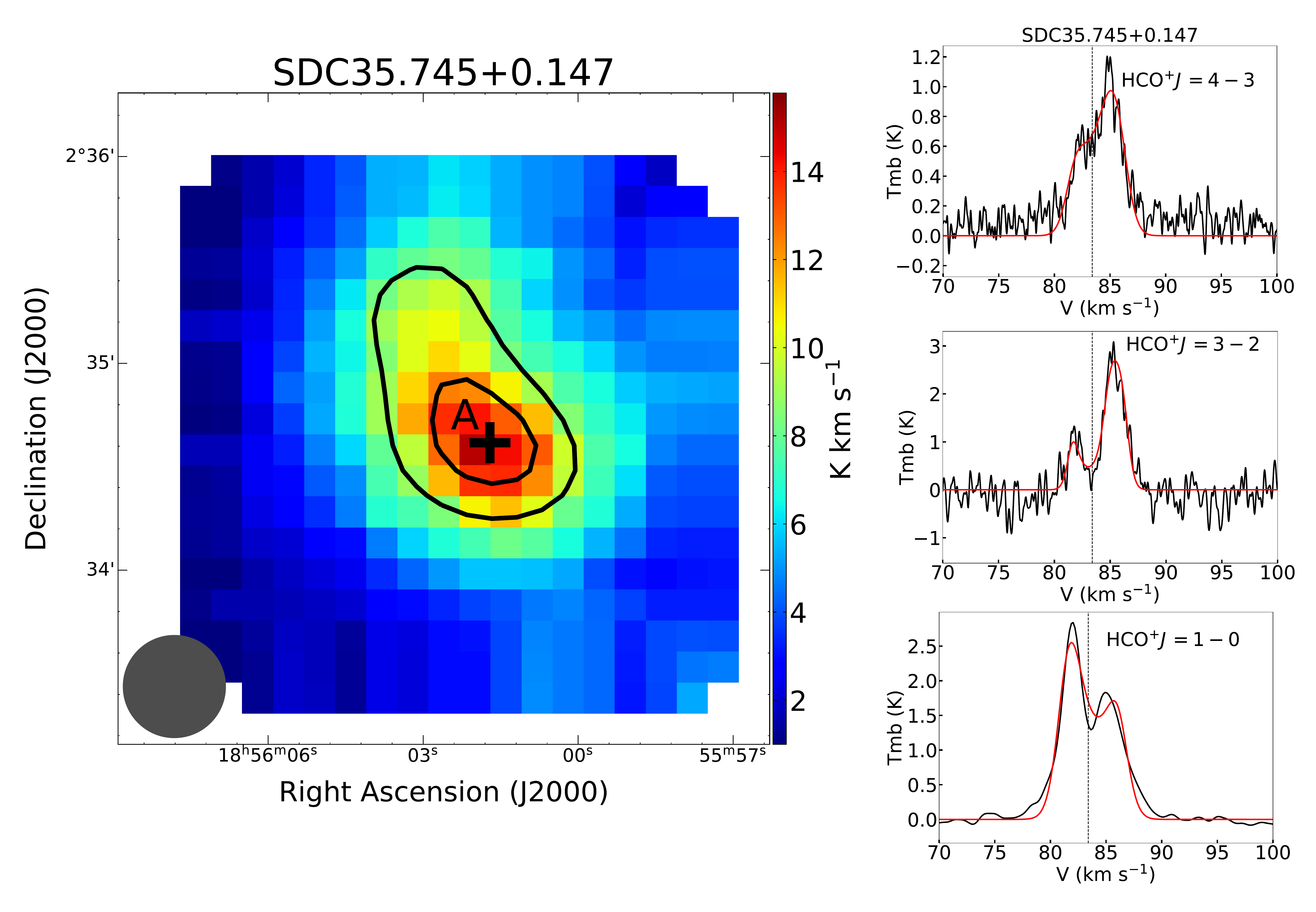} 
    \caption{The integrated intensity map of {\HCOn}$J\,$=1--0 overlaid with {\HCOn}$J\,$=4--3 contours at 0.5, 0.8, and 1.0 of the maximum of the integrated intensity in K {\kms} for SDC35.745. The maximum of the integrated intensity has been stated in Table~\ref{tab:size_luminosity} as the biggest value of $\int T_{MB}dv$ for the source. The grey dot at the corner of the emission map represents the IRAM beam size.}
    \label{fig:sdc35p745}
\end{figure}

We applied the HILL5 model to fit {\HCOn}$J\,$=1--0, 3--2, and 4-3 spectral lines towards the peak positions as listed in Table~\ref{tab:infall_velocity}. Fitted lines are plotted in red in Figure~\ref{fig:sdc18p624} through Figure~\ref{fig:sdc35p745}. The fitted infall velocities are listed in Table \ref{tab:infall_velocity}. We exclude clumps (SDC18.888-0.476A, SDC24.618, and SDC35.429) which have red-asymmetries for all transitions.

\begin{table}
\scriptsize
\rotatebox{90}{\begin{minipage}[]{120mm}\caption[]{Infall velocities fitted by HILL at Peak Positions of {\HCOn}$J\,$=4--3 \label{tab:infall_velocity}}\end{minipage}}
\centering
\rotatebox{90}{
\setlength{\tabcolsep}{2.5pt}
 \begin{tabular}{lcc c c c c c c c c c }
  \hline\noalign{\smallskip}
Source Name & R.A.(J2000) & Decl.(J2000) & & {\HCOn}$J\,$=1--0    & &   & {\HCOn}$J\,$=3--2 & & & {\HCOn}$J\,$=4--3 & \\
 & (hh:mm:ss) & ($^\circ$ ' ") & V$_{in}$ ({\kms})  & $\sigma$ ({\kms})  & $\tau$&  V$_{in}$ ({\kms}) & $\sigma$ ({\kms})  & $\tau$  & V$_{in}$ ({\kms})  & $\sigma$ ({\kms})  & $\tau$ \\
  \hline\noalign{\smallskip}
SDC18.624-0.070A & 18:25:10.8 & -12:42:20 &    $0.85^{+0.19}_{-0.16}$   &$1.42^{+0.09}_{-0.08}$  &$2.60^{+0.36}_{-0.35}$  & $0.31^{+0.11}_{-0.13}$ & $1.36^{+0.03}_{-0.03}$ & $0.96^{0.06}_{0.03}$ &  - & - & -\\
SDC18.624-0.070B & 18:25:08.5 & -12:45:25 & $1.33^{+0.05}_{-0.06}$  & $0.82^{+0.03}_{-0.02}$  & $7.99^{+0.01}_{-0.18}$ & $1.47^{+0.15}_{-0.15}$ & $1.21^{+0.10}_{-0.09}$ & $4.21^{+0.59}_{-0.49}$ & $0.74^{+0.15}_{-0.26}$ & $1.16^{+0.21}_{-0.13}$ & $0.79^{+0.41}_{-0.20}$  \\
SDC18.888-0.476A &18:27:09.7 & -12:41:32 & -& - & - & - & - & -  & - & - & -\\ 
SDC18.888-0.476B &18:27:07.1 & -12:41:40& $0.57^{+0.21}_{-0.28}$& $1.87^{+0.07}_{-0.09}$ & $0.53^{+0.03}_{-0.02}$& $1.88^{+0.11}_{-0.11}$& $2.00^{+0.01}_{-0.50}$& $1.21^{+0.05}_{-0.05}$& $0.85^{+0.10}_{-0.27}$& $2.33^{+0.10}_{-0.03}$ & $0.35^{+0.02}_{-0.02}$ \\ 
SDC22.373+0.446 & 18:30:24.5 & -09:10:34 &$0.97^{+0.10}_{-0.10}$ & $0.84^{+0.05}_{-0.05}$ & $2.88^{+0.35}_{-0.33}$  & $0.94^{+0.13}_{-0.15}$ & $0.75^{+0.08}_{-0.07}$ & $4.63^{+1.08}_{-0.85}$ & $0.77^{+0.15}_{-0.12}$  & $0.89^{+0.52}_{-0.66}$ & $2.09^{+0.30}_{-0.32}$\\
SDC23.367-0.288A & 18:34:53.9 & -08:38:22  & $1.16^{+0.34}_{-0.31}$ & $2.45^{+0.12}_{-0.23}$ & $0.50^{+0.15}_{-0.05}$ & $1.47^{+0.37}_{-0.43}$ & $2.00^{+0.06}_{-0.59}$ & $0.32^{+0.04}_{-0.03}$  & $0.81^{+0.20}_{-0.44}$  & $1.15^{+0.21}_{-0.13}$ & $0.81^{+0.49}_{0.56}$   \\
SDC23.367-0.288B & 18:34:52.4 & -08:36:47  & $1.35^{+0.40}_{-0.42}$ & $1.13^{+0.20}_{-0.21}$ & $3.17^{+1.22}_{-0.99}$ & $1.30^{+0.71}_{-0.50}$ & $1.45^{+0.23}_{-0.31}$ & $2.40^{+0.93}_{-0.98}$ & $1.32^{+0.20}_{-0.77}$  & $0.97^{+0.22}_{-0.11}$ & $2.10^{+1.04}_{-0.90}$\\
SDC24.489-0.689A & 18:38:25.8 & -07:49:36  & $0.98^{+0.10}_{-0.10}$ & $1.04^{+0.05}_{-0.05}$& $4.83^{+0.52}_{-0.47}$  & $0.49^{+0.21}_{-0.16}$ &$0.97^{+0.08}_{-0.07}$ & $5.09^{+1.19}_{-1.00}$ & $0.80^{+0.38}_{-0.26}$  & $1.28^{+0.13}_{-0.17}$ & $1.40^{+0.40}_{-0.56}$ \\
SDC24.489-0.689B & 18:38:28.4 & -07:49:05  & $1.85^{+0.14}_{-0.13}$  & $1.24^{+0.09}_{-0.08}$ & $5.37^{+0.90}_{-0.72}$ & -&  - & -  & -  & - & - \\
SDC24.618-0.323A & 18:37:22.9 & -07:31:42 &-  & - & -& -& - & - & -  & - &-\\
SDC24.618-0.323B &18:37:21.5 & -07:33:20 & -  & - & - & - &- & - &- & - & -\\
SDC25.166-0.306A & 18:38:09.7 & -07:02:31  & $0.79^{+0.09}_{-0.08}$& $1.20^{+0.04}_{-0.04}$ & $3.68^{+0.30}_{-0.28}$ &-  & -& - & - &- & - \\
SDC25.166-0.306B & 18:38:18.1 & -07:02:52  & $0.61^{+0.09}_{-0.08}$ & $1.05^{+0.05}_{-0.05}$ & $5.09^{+0.57}_{-0.52}$ & $0.56^{+0.14}_{-0.12}$ & $0.82^{+0.06}_{-0.06}$ & $5.67^{+1.39}_{-1.11}$ & $0.60^{+0.20}_{-0.23}$ & $0.92^{+0.15}_{-0.12}$ & $1.09^{+0.50}_{-0.30}$ \\
SDC25.166-0.306C & 18:38:13.1 & -07:03:08  & $0.88^{+0.12}_{-0.11}$  & $1.10^{+0.05}_{-0.05}$ & $4.24^{+0.47}_{-0.44}$ & $0.68^{+0.73}_{-0.30}$  & $0.88^{+0.16}_{-0.27}$ & $3.87^{+1.87}_{-1.97}$  & $0.49^{+0.13}_{-0.30}$ & $0.97^{+0.07}_{-0.13}$ & $0.80^{+0.16}_{-0.10}$ \\
SDC28.333+0.063A & 18:42:50.7 & -04:03:14  & $0.45^{+0.11}_{-0.16}$ & $1.00^{+0.01}_{-0.01}$ & $0.80^{+0.03}_{-0.03}$ & $1.47^{+0.25}_{-0.23}$  & $1.44^{+0.11}_{-0.12}$ & $6.38^{+2.01}_{-1.46}$ & $0.79^{+0.13}_{-0.14}$  &  $1.23^{+0.08}_{-0.07}$ & $1.02^{+0.20}_{-0.20}$ \\
SDC28.333+0.063B & 18:42:52.2 & -03:59:56 & - & - & - & $1.52^{+0.34}_{-0.26}$  & $1.43^{+0.15}_{-0.14}$ & $3.99^{+0.87}_{-0.76}$ & $0.37^{+0.07}_{-0.07}$ & $1.14^{+0.04}_{-0.04}$& $5.38^{+0.55}_{-0.50}$ \\
SDC28.333+0.063C & 18:42:54.1 & -04:02:30 & $0.77^{+0.32}_{-0.28}$ & $2.92^{+0.20}_{-0.17}$ & $5.54^{+1.09}_{-0.83}$ & $0.09^{+0.34}_{-0.26}$ & $0.74^{+0.13}_{-0.09}$ & $2.03^{+0.98}_{-0.93}$ & $0.34^{+0.13}_{-0.18}$ & $1.20^{+0.04}_{-0.09}$ & $0.80^{+0.04}_{-0.02}$ \\
SDC28.333+0.063D & 18:42:49.3 & -04:02:17 & $2.70^{+0.33}_{-0.31}$  &$2.20^{+0.23}_{-0.22}$ & $7.09^{+2.24}_{-1.52}$ & $1.31^{+0.38}_{-0.33}$ & $1.20^{+0.15}_{-0.15}$ & $4.49^{+1.47}_{-1.16}$ & $0.46^{+0.32}_{-0.77}$  & $1.45^{+0.12}_{-0.18}$ & $0.30^{+0.23}_{-0.02}$ \\
SDC35.429+0.138 & 18:55:34.1 & +02:19:07  &- &- & -& $0.70^{+0.16}_{-0.60}$ & $1.18^{+0.23}_{-0.14}$ & $0.47^{+0.40}_{-0.20}$ & -  & - & - \\
SDC35.527-0.269A & 18:57:09.1 & +02:07:55  & $1.03^{+0.14}_{-0.16}$ & $1.35^{+0.13}_{-0.10}$ & $1.29^{+0.32}_{-0.35}$ & $0.87^{+0.12}_{-0.01}$ &$0.62^{+0.01}_{-0.09}$& $1.00^{+0.10}_{-0.07}$ & $0.32^{+0.13}_{-0.47}$ & $0.81^{+0.05}_{-0.13}$ & $0.80^{+0.35}_{-0.20}$  \\
SDC35.527-0.269B & 18:57:08.0 & +02:10:56 & $0.56^{+0.26}_{-0.29}$  & $1.01^{+0.11}_{-0.12}$  & $0.83^{+0.31}_{-0.49}$ & $0.54^{+0.15}_{-0.51}$ & $0.60^{+0.26}_{-0.15}$  & $0.98^{+0.01}_{-0.85}$ & $0.22^{+0.41}_{-0.82}$  & $1.03^{+0.13}_{-0.25}$ & $0.10^{+0.51}_{-0.35}$ \\
SDC35.527-0.269C & 18:57:08.6 & +02:09:07 & $0.90^{+0.30}_{-0.47}$ & $1.19^{+0.24}_{-0.19}$&  $1.28^{+0.63}_{-0.60}$& - &- & -& $0.51^{+0.40}_{-0.06}$ & $1.11^{+0.18}_{-0.24}$ & $0.80^{+0.45}_{-0.10}$ \\
SDC35.745+0.147 &18:56:01.7 & +02:34:37 & $0.54^{+0.08}_{-0.08}$ & $1.36^{+0.05}_{-0.05}$ & $3.06^{+0.26}_{-0.25}$ & -& -&-& $-0.85^{+0.17}_{-0.20}$ &$1.12^{+0.10}_{-0.10}$ &  $1.82^{+0.41}_{-0.42}$ \\
\hline
  \noalign{\smallskip}\hline
\end{tabular}
}
\rotatebox{90}{
\tablecomments{1.3\textwidth}{The upper and lower limits of infall velocities are determined within 3$\sigma$ confidence interval limit. We excluded the values from the profiles where there is no infall signature.}
}
%% place \tablecomments and \tablerefs below \end{center| and \end{center}:
%% you may leave the table-width parameter to editors or set to your actual size
%\tablecomments{0.86\textwidth}{The SDSS $ugriz$ magnitudes are given in AB system 
%and the UKIDSS $YJHK$ magnitudes are given in Vega system.}
\end{table}

In some sources (e.g. SDC25.166-0.306C), although the blue-asymmetric profile look differently for different transitions, the derived infall velocities are similar ($\Delta V_{in} <$ 0.2\,{\kms}). A possible explanation is that the line has different optical depth, as listed in Table~\ref{tab:infall_velocity}, which causes the differences in the line profiles, even with similar infall velocities.

\subsection{Clump Sizes and Line Luminosities from {\HCOn}$J\,$=4--3 maps}

{\HCOn}$J\,$=4--3 traces the densest part of the cloud among the three transitions, thus can be used to define the compact, very dense regions in these IRDCs. We obtained the effective angular diameter ($\theta_{transition}$) size of the dense clumps by measuring the area of the contour at the intensity half of the peak intensity of {\HCOn}$J\,$=4--3, following the method used in \citet{2010wu}, with the equation: 

\begin{equation}\small
\begin{aligned}
\theta_{transition} = 2\left(\frac{A_{1/2}}{\pi}-\frac{\theta^2_{beam}}{4}\right)^{1/2},\label{eq:clump_size}
\end{aligned}
\end{equation}
where $A_{1/2}$ is the area within the contour of half-peak intensity and $\theta_{beam}$ is the angular beam size. $R_{transition}$ is the equivalent spatial radius of the clumps, and is calculated as $R_{transition} = \theta_{transition} D/2$, where $D$ is the distance of the source.

The line luminosity is then calculated as: 
\begin{equation}\small
\begin{aligned}
L' = 23.5 \times 10^{-6} \times D^{2} \times \left(\frac{\pi \times \theta^2_{transition}}{4 ln 2} \right) \times \left(\frac{\theta^2_{transition}+\theta^{2}_{beam}}{\theta^2_{transition}}\right) \times \int T_{mb\,} \textrm{dv K km s}^{-1},
\label{eq:line_luminosity}
\end{aligned}
\end{equation}
where $T_{mb}dv$ is the integrated intensity at the peak position, $D$ is in kpc and $\theta$ in arcsecond \citep{2010wu}. The calculated clump angular sizes, spacial sizes $R$, and the line luminosities for {\HCOn}$J\,$=4--3 are listed in Table~\ref{tab:size_luminosity}. 

These IRDC clumps have {\HCOn}$J\,=$4--3 sizes ranging from 0.17\,pc to 0.73\,pc, with a mean size of 0.33 pc and a median size of 0.28 pc. These sizes are comparable to clump sizes of higher-$J$ transitions found in later stages massive clumps. For example, The half-peak clumps' sizes of HCN$J\,$=3--2 towards some UCHII clumps \citep{2010wu} have a mean and median size of 0.32\,pc and 0.26\,pc, respectively. The derived line luminosity of {\HCOn}$J\,=$4--3 ranges 0.2-18\,K {\kms} pc$^{2}$, with a mean and median value of 4.1\,K {\kms} pc$^{2}$ and 2.6\,K {\kms} pc$^{2}$, respectively.

\begin{table}[h!]
\bc
\begin{minipage}[]{150mm}
\caption[]{The derived parameters of IRDC clumps. \label{tab:size_luminosity}}\end{minipage}
\setlength{\tabcolsep}{2.5pt}
\small
 \begin{tabular}{lcc c c c c c  }
  \hline\noalign{\smallskip}
Source Name & Clump Size$^{a}$  & $R_{transition}$ $^{a}$ & Line Luminosity\add{$^{a}$}  & $\int T_{MB}dv$\add{$^{a}$}  & $\overline{N_{H_{2}}}^{b}$ & $M_{H_{2}}^{b}$ & $\dot M^{c}$ \\
 & (")& (pc) & (K {\kms} pc$^{2}$)  & (K {\kms})  & $10^{22}\,cm^{-2}$  & ({\Msun}) & ($\times 10^{-3}$\,{\unitMsunyr})\\
  \hline\noalign{\smallskip}
SDC18.624-0.070A & 31 & 0.26 &  2.74 & 3.97 & - & - &- \\
SDC18.624-0.070B & 33 & 0.28 & 3.05  & 6.76  & 4.2 & 636 & 10.3\\
SDC18.888-0.476A & -  & - &  - & 8.43 & -& - &-\\ 
SDC18.888-0.476B & 49  & 0.52 &  17.64  & 12.02  & 24.5 & 3430 & 28.4 \\ 
SDC22.373+0.446 & 26 & 0.23  &  1.22 & 3.51 & 6.7 & 564 & 4.8 \\
SDC23.367-0.288A & 25 & 0.28  &  1.16 & 2.02 & 12.9& 1597 & 10.6\\
SDC23.367-0.288B & 15  & 0.17  &  0.42 & 1.46& 1.1 & 288 & 1.2 \\
SDC24.489-0.689A &51  & 0.41 & 4.13  & 4.35 & 8.3 & 426& 4.4\\
SDC24.489-0.689B &-  & - & - & 2.41 &-  & - &- \\
SDC24.618-0.323A &37  &0.27 & 2.35 & 5.24 & -& - & - \\
SDC24.618-0.323B & -   & -  & - & 2.16& -&-& -\\
SDC25.166-0.306A & 47  & 0.45 & 4.32  & 3.00 & - & -& -\\
SDC25.166-0.306B & - & - & - & 1.95 & 4.9 & 734 & 2.1\\
SDC25.166-0.306C & - & - & - & 1.56 & 5.1& 830 & 3.1\\
SDC28.333+0.063A & 25 & 0.28 & 2.04 & 4.56 & 5.4& 420 & 4.8 \\
SDC28.333+0.063B & 35 & 0.39 & 5.90 & 6.99& - & - & -\\
SDC28.333+0.063C & - & - & - & 3.09  & 4.1 & 291& 5.8\\
SDC28.333+0.063D & - & - & - & 3.16 & 3.7 & 275 & 19.8\\
SDC35.429+0.138 & 37 & 0.42 & 8.47 & 6.67 & - & -& - \\
SDC35.527-0.269A & 29 & 0.21 & 0.51& 2.08 & 3.0& 280 & 1.8 \\
SDC35.527-0.269B &28 & 0.20 &0.49& 1.59 & 2.4& 248 & 0.9\\
SDC35.527-0.269C &28 & 0.20& 0.23 & 0.91& -  & - &-  \\
SDC35.745+0.147 & 59 & 0.73 & 11.50 & 3.79 & - & - & -\\
\hline
Mean & 35 & 0.33 & 4.13 & 3.99& 6.6 & 770 & 7.5 \\
Median & 32 & 0.28 & 2.55 & 3.13 & 5.0 & 490&4.8\\
\hline
  \noalign{\smallskip}\hline
\end{tabular}
\tablecomments{0.98\textwidth}{$^{a}$The physical parameters are from {\HCOn} $J\,$=4--3 observations. The clump size and $R_{transition}$ are determined from half peak intensity of {\HCOn} $J\,$=4--3 observations. For some sources, the sizes of half intensity contours are smaller than the beam size that the sources are unresolved. Thus, we can not calculate their sizes and masses. $^{b}$The physical parameters are from {\it Herschel} 160\,$\mu$m and 250\,$\mu$m observations, see sect. \ref{subsect:mass_acc_rate} for details. $^{c}$ The physical parameters are derived from {\HCOn} $J\,$=4--3 observations, see sect. \ref{subsect:mass_acc_rate} for details.}
\ec
\end{table}

\subsection{Clump Mass and Mass Accretion Rate}\label{subsect:mass_acc_rate}

 Assuming all the mass within the beam is uniformly distributed and contributing to the infall motions and the peak regions are spherical, following \citet{2018calahan} the mass accretion rate is calculated as:
 
\begin{equation}
\begin{aligned}
{\dot M} & =4 \pi R^{2} \rho v_{in}=\frac{3Mv_{in}}{R} \\
&=3068 {\frac{M_{\odot}}{Myr}}{\left(\frac{M}{1000\,M_{\odot}}\right)} \left(\frac{v_{in}}{1\,km s^{-1}}\right) \left(\frac{1\,pc}{R}\right),
\label{eq:mass_accretion_rate}
\end{aligned}
\end{equation}

The infall velocities used to calculate the mass accretion rates for these IRDCs are from the modelling of {\HCOn}$J\,$=1--0 observations. We use the column density maps from \citet{2016peretto} to calculate the mass then the average volume density of the clumps within the IRAM {\HCOn}$J\,$=1--0 beam to match the radius from infall observations, to estimate the mass accretion rates. \citet{2016peretto} derived the column density maps of the sources from {\it Herschel} 160\,$\mu$m and 250\,$\mu$m observations obtained as part of the Hi-GAL survey \citep{2016molinari} and they estimate an uncertainty of $\sim$50\% in the column density \citep{2016peretto}. 

The assumption that all mass within the beam will contribute to infall profile could lead to an overestimate of the averaged volume density within the beam due to the foreground and background emission along the light of sight, therefore an overestimate on the mass accretion rate.
The averaged column density within the beam, as well as the derived mass accretion rate of the sources are listed in Table~\ref{tab:size_luminosity}. 

Note that the above mass estimated from the column density map within the IRAM beam is a beam-size defined mass.
Here we want to estimate the physical mass of these clumps by calculating the total dust masses above $4\sigma$ of the {\it Herschel} map. Based on the radiative transfer equation, following \citet{1983Hildebrand}, the flux density of the dust core is related to its mass in the form of 
\begin{equation}\label{equ:core_mass}
\begin{aligned}
S_{\nu} & = \kappa_\nu B_{\nu}(T_{\rm d}) \Omega \mu m_{\rm H} N_{\rm tot} \\
        & = \frac{\kappa_{\nu} B_{\nu}(T_{\rm d}) M_{\Omega} }{D^2},
\end{aligned}
\end{equation}
wherein $S_{\nu}$ is the flux density at the frequency $\nu$. $\Omega$ is the solid angle of the core or selected area. $B_{\nu}(T_{\rm d})$ is the Planck function of the dust temperature $T_{\rm d}$, $N_{\rm tot}$ is the gas column density (mostly HI+H$_2$). $\mu=2.33$ is the average molecular weight, $m_{\rm H}$ is the mass of the hydrogen atom, and $\kappa_\nu$ is the dust opacity, which is assumed to be related to the frequency in the form $\kappa_\nu=\kappa_{\rm 230 GHz}(\nu/{\rm 230\,GHz})^{\beta}$. The reference value $\kappa_{\rm 230 GHz}=0.009$ cm$^2$ g$^{-1}$, is adopted from dust model for the grains with coagulation for $10^5$\,yr with accreted ice mantles at a density of $10^6$\,cm$^{-3}$ \citep{1994Ossenkopf}. $D$ is the source distance. 
In calculation, we used the NH$_3$ kinetic temperature (Xie et al. 2021 in press) to approximate the dust temperature. The total core mass which covers the regions above $4\sigma$ in the dust emission map have been calculated using equation \ref{equ:core_mass}, and the result is shown in Table \ref{tab:size_luminosity}. The derived masses of these IRDC clumps range from 250 to 3400 M$_{\odot}$, with a mean mass of 770 M$_{\odot}$ and a median mass of 490 M$_{\odot}$. They are massive enough to form massive star clusters.

The average and mean value of the mass accretion rates are 7.5$\times10^{-3}$\,{\unitMsunyr} and 4.8$\times10^{-3}$\,{\unitMsunyr}, respectively. These values are slightly larger than those found in later massive star forming regions like HMPOs or UCHII regions, as seen in Table~\ref{tab:previous_studies}. Given that the infall velocity derived from the two-layer model will be underestimated by a factor of 2 compared to HILL5 model, the mass accretion rates in these IRDCs are not very different to those in HMPOs or UCHII regions. 

The Kelvin-Helmholtz timescale, which is the time it takes for a star to radiate away its total kinetic energy, reflects how long a star can sustain its luminosity by gravitational contraction. An empirical form of Kelvin-Helmholtz timescale $t_{KH}$ can be written as \citep{2002PhDT........10L}:
\begin{equation}
\begin{aligned}
t_{KH} \approx 19 Myr \left(\frac{M_{*}}{M_{\odot}}\right)^{-2.5},
\label{eq:kh_timescale_sim}
\end{aligned}
\end{equation}
where M$_{*}$ is the stellar mass. For a star sufficiently massive, $t_{KH}$ can be too short for a star to even reach the main sequence \citep{2003li,2007klaassenwilson}. For example, an 8\,{\Msun} stellar mass, the lower end of massive stars, has a Kelvin-Helmholtz timescale of 0.1\,Myr. The mass accretion rate has to be high enough to allow mass accumulated within 0.1\,Myr to be more than 8\,{\Msun}, so that the massive star will survive to main-sequence. Our measured accretion rate of 5$\times10^{-3}$\,{\unitMsunyr} in these IRDC clumps corresponds to the infall of $\sim$500\,{\Msun} in 0.1\,Myr. According to equation Eq.~\ref{eq:kh_timescale_sim}, for any stellar mass M$_{*}$ $\lesssim$ 26\,{\Msun}, the current infall rate can provide enough accretion within its Kelvin-Helmholtz time, but more massive stars will be hard to reach main-sequence. Therefore, these IRDC clumps can form massive stars, but are more likely to form star clusters rather than a single massive star.

\section{Discussion}
\label{sect:discussion}

\subsection{Infall tracers for IRDCs}

Some works have claimed that {\HCOn}$J\,$=4--3 is likely more sensitive to trace infall than {\HCOn}$J\,$=1--0 in massive star forming regions, especially for denser regions \citep{2008tsamis, 2006purcell, 2012klaassen}. While in a survey towards a large sample of HMPOs, {\HCOn}$J\,$=1--0 has been found to have more prominent infall signatures, i.e. double peaks with self-absorbed dip, than {\HCOn}$J\,$=4--3 lines \citep{2005fuller}, suggesting lower-$J$ {\HCOn} transitions could be a better infall tracer in  these regions. Very limited such studies have been done on IRDCs \citep[e.g.][]{2016wyrowski}. Based on our survey with multiple {\HCOn} lines towards a sample of 23 clumps, we can evaluate their ability in revealing infall in IRDCs.

\begin{figure}
    \includegraphics[width=.90\textwidth]{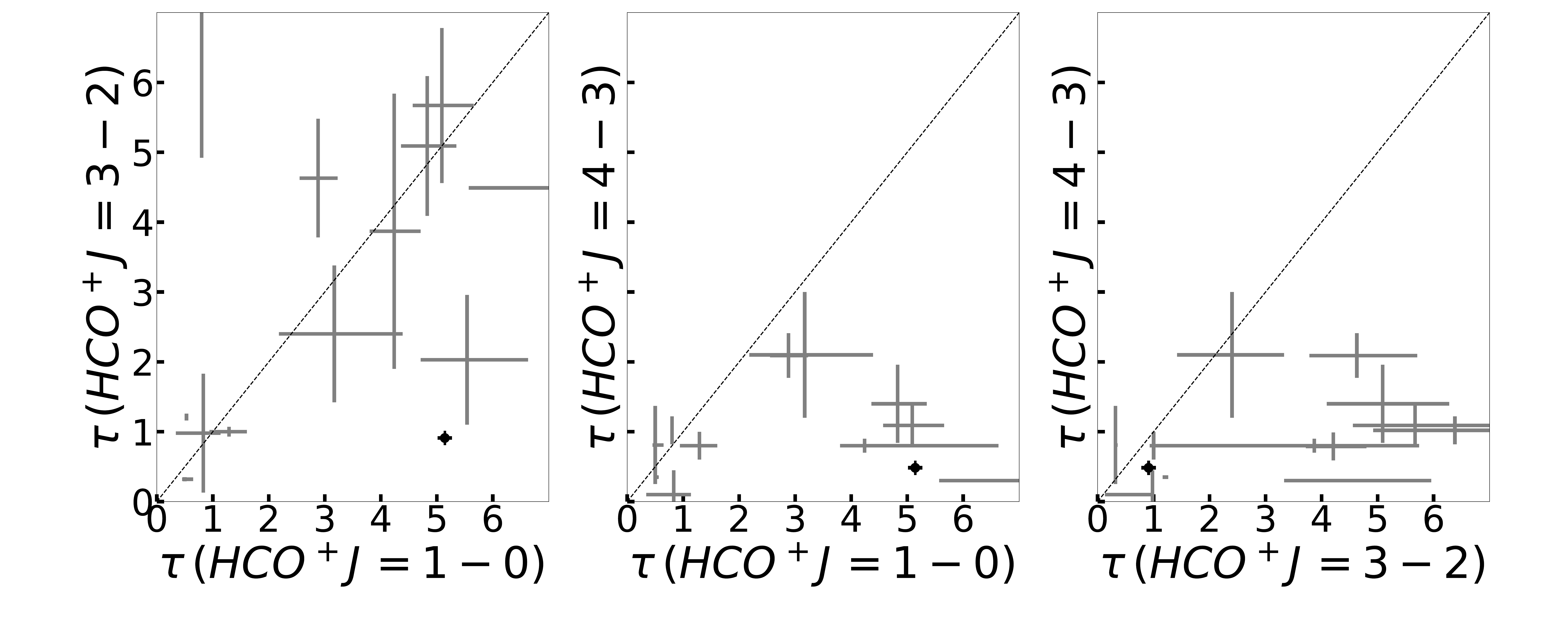}
    \caption{The optical depth comparisons among the three transitions: $J\,$=1--0, $J\,$=3--2, and $J\,$=4--3. The diagonal line indicates where the optical depths are equal.  The black markers indicate the weighted average of optical depths. Despite the scattering of the optical depths of {\HCOn}$J\,$=3--2 and {\HCOn}$J\,$=1--0, they are similar and both are higher than those of {\HCOn}$J\,$=4--3. }
    \label{fig:optical_depth}
\end{figure}

As seen from Table~\ref{tab:number_profile}, the surveys with three {\HCOn} transitions at least imply that in infall candidates selected by {\HCOn}$ J\,=$1--0, {\HCOn}$J\,$=4--3 is not more sensitive, not even comparable, in detecting infall in IRDCs. These results suggest {\HCOn}$ J\,=$4--3 may not be optically thick enough to reveal infall in these IRDCs, as also indicated by the opacity derived from HILL model in Table~\ref{tab:infall_velocity} and Figure~\ref{fig:optical_depth}.

\begin{figure}
    \includegraphics[width=.90\textwidth]{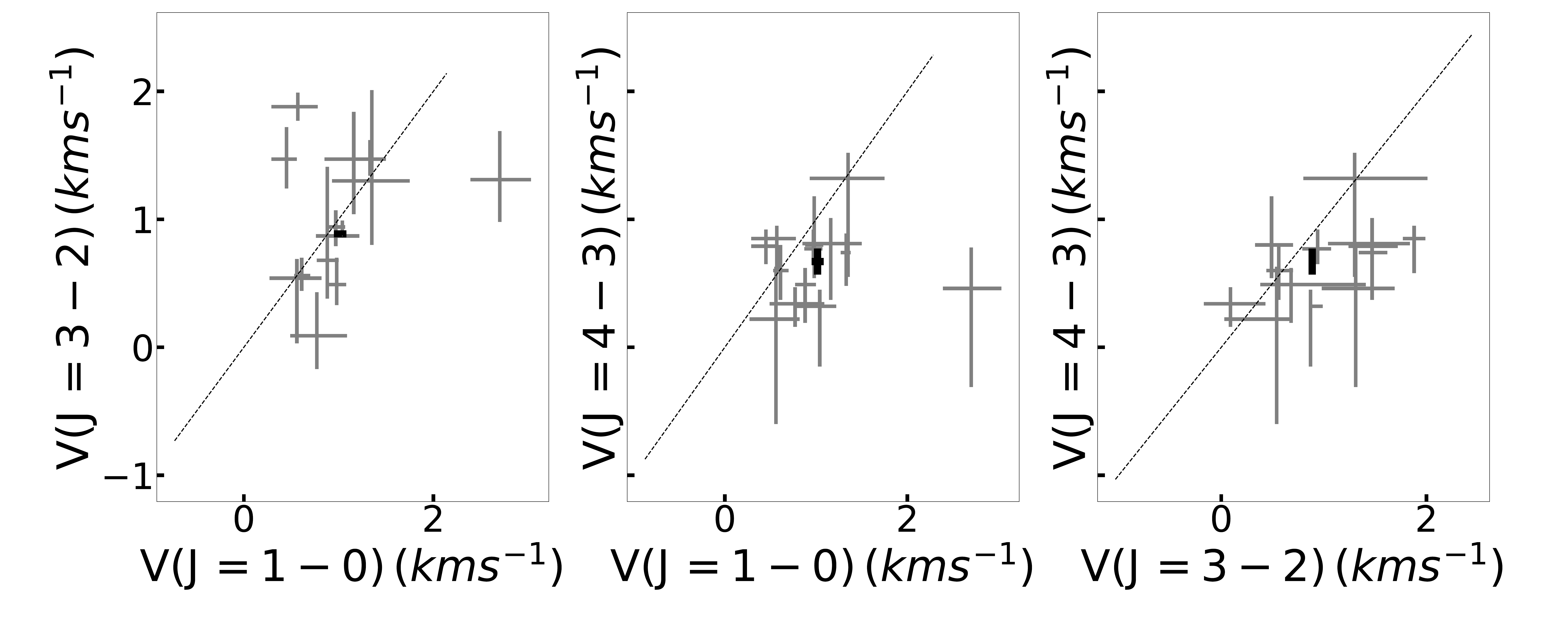}
    \caption{The infall velocity comparisons among the three transitions: $J\,$=1--0, $J\,$=3--2, and $J\,$=4--3. The diagonal line indicates where the infall velocities are equal. The black markers indicate the weighted average of infall velocities.}
    \label{fig:infall_velocity_comparison}
\end{figure}

{\HCOn}$J\,$=3--2 has been found to have comparable ability to {\HCOn}$J\,$=1--0 to study infall in later stages of massive star forming regions like HMPOs \citep{2005fuller,2011reiter,2012klaassen}, and we find a similar trend for {\HCOn}$J\,$=3--2 in IRDCs. As well as the line profiles, we also compare the derived infall velocities of the three {\HCOn} transitions as presented in Table~\ref{tab:infall_velocity}. The infall velocities derived from {\HCOn}$J\,$=1--0 are slightly larger than those from {\HCOn}$J\,$=3--2 overall, while the weighted average of both transitions are similar (1.02$\pm$0.03\,{\kms} for {\HCOn}$J\,$=1--0 and 0.89$\pm$0.01\,{\kms} for {\HCOn}$J\,$=3--2). This may suggest that throughout the parts or layers in the clump where {\HCOn}$J\,$=1--0 and 3--2 probe, the infall velocity does not change much, agreeing with the hypothesis that infall velocity is consistent within these layers or regions. The infall velocities of {\HCOn}$J\,$=4--3, on the other hand, are much smaller. The weighted average infall velocity of {\HCOn}$J\,$=4--3 (0.67$\pm$0.05\,{\kms}) is more than 1.5 times smaller than what is derived from {\HCOn}$J\,$=1--0, with a similar result been found with another study using the two-layer model towards an IRAS source \citep{2010barnes}. Excluding the non-fitted and/or the negative values in Table~\ref{tab:infall_velocity}, the median values of infall velocities for {\HCOn}$J\,$=1--0, 3--2, and 4--3 are 0.97, 0.94, and 0.74\,{\kms}, respectively. The comparisons between pairs of transitions of {\HCOn} are shown in Figure~\ref{fig:infall_velocity_comparison}, which suggests that the infall velocity of {\HCOn}$J\,$=3--2 is comparable to that of {\HCOn}$J\,$=1--0, while both are larger than those of {\HCOn}$J\,$=4--3. The opacities derived from the HILL5 model for all three transitions are also listed in Table~\ref{tab:infall_velocity}, and are compared between every two transitions for each clumps in Figure~\ref{fig:optical_depth}. It is clear to see that the opacities calculated from {\HCOn}$J\,$=1--0 and 3--2 are comparable and both are much higher than the values from  {\HCOn}$J\,$=4--3. This is consistent with RADEX \citep{radex} modelling for 10$^{4}$--10$^{5}$\,cm$^{-3}$ at temperatures from 15 to 30\,K. The detailed modelling is beyond the scope of this observational work and will be presented in future works.  

\subsection{Infall properties of IRDCs}

Different infall tracers, especially lower-$J$ and higher-$J$ {\HCOn} or HCN transitions normally trace different parts or layers of dense star forming clumps. For massive clumps at different evolutionary stages, like UCHII regions, HMPOs and IRDCs, we will need different infall tracers to best probe the collapsing regions in the clumps, as argued in \citet{2003wu} that infall profile will best be revealed when the opacity of the source and the critical density of the tracer can be well matched. Therefore, the infall properties may be compared for samples at different evolutionary stage as long as the best infall tracers are used to model the kinematic information of infall.

Under this assumption, we find that the infall velocities of IRDCs obtained in this work are in general consistent with the values in other massive star forming regions. For example, an infall velocity of $\sim$1.5\,{\kms} for {\HCOn}$J\,$=3--2 was obtained using HILL5 model towards an IRDC \citep{2018contreras} with mass approximate to the mass of SDC23.367-0.288, which has an average infall velocity of $\sim$1.4\,{\kms}. In Table~\ref{tab:previous_studies}, we listed infall velocities as well as mass accretion rates obtained from some other studies towards HMPOs and UCHII regions, to compare with our results. Considering infall velocities calculated from two-layer model are generally a factor of 2 less than obtained from the HILL model, the derived infall velocity and therefore the mass accretion rate of these IRDCs (with a median of 1.0\,{\kms}) is actually comparable to those calculated from HMPOs \citep{2005fuller} or UCHII regions \citep{2007klaassenwilson}, which is interesting as this suggests that although these molecular line tracers may origin from different parts/layers of the clumps, they imply that infall velocities are within a similar range for these different samples.

\begin{table}
\bc
\begin{minipage}[]{90mm}
\caption[]{Previous Infall Studies.\label{tab:previous_studies}}\end{minipage}
\setlength{\tabcolsep}{1.0pt}
\small
 \begin{tabular}{l  c c c  c c c }
  \hline\noalign{\smallskip}
Targets $^{a}$  & Tracer & Method & V$_{in}$ Range ({\kms}) & Mean V$_{in}$ ({\kms}) & $\dot{M}$ Range ({$\times 10^{-3}$\unitMsunyr}) & Reference\\
  \hline\noalign{\smallskip}
 HMPOs (22/77)& (1--0)  & two-layer & [0.1, 1.0] & -   & [0.2, 1.0] & (1)\\
 UCHII (9/23) & (4--3) & two-layer & [0.1, 1.8] & $\sim$0.9  & [0.02, 10]  & (2) \\
UCHII (8/30) &(3--2) & two-layer & [0.1, 1.3]& $\sim$0.5  & [0.03, 6] & (3)\\
BGPS clump (6/101) & (1--0)  & HILL5 & [0.3, 0.8]&  $\sim$0.7 & [0.5, 2] & (4)\\
 Hi-GAL clump (21/213) & (1--0)  & two-layer & [0.2, 1.5] & $\sim$0.3  & [0.7, 45.8]  & (5)\\
 four IRDCs & (1--0) & two-layer & [0.4, 2.4] & $\sim$1.0 &  [2.0, 18.0] & (6)\\
\hline
IRDCs  & (1--0) & HILL5 & [0.5, 2.7] & $\sim$1.0 & [2.0, 17.2] & This Work \\
\hline
  \noalign{\smallskip}\hline
\end{tabular}
\tablecomments{1.00\textwidth}{The references are: (1) \citet{2005fuller}; (2) \citet{2007klaassenwilson}; (3) \citet{2010churchwell}; (4) \citet{2018calahan}; (5) \citet{2018traficante}; (6) \citet{2015giannetti}. $^{a}$ The numbers in the parentheses are the numbers of infall candidates versus the total numbers of the sources. The infall velocity derived from two-layer model has been found about a factor of 2 smaller than calculated from the HILL5 model \citep{2005hill, 2010barnes}. }
\ec
\end{table}

\subsection{Infall velocities and the {\NHn} to CCS ratio}
 Some works have argued that the ratio of the column densities of {\NHn} and CCS could be an indicator of chemical evolution in low mass and high mass star formation \citep[e.g.][]{1992suzuki,2008sakai}. Should this ratio be an indicator of chemical evolution within the IRDC stage in massive star formation, we have tested to see if there is any correlation between the obtained infall velocity and the logarithm of the ratio of the column density of {\NHn} to that of CCS. To best match the resolution of the {\NHn} and CCS survey (HPBW$\sim$45") (Xie et al. 2021, in press), we use the infall velocities derived from {\HCOn}$J\,$=1--0 (HPBW$\sim$30"). The results are shown in Figure \ref{fig:evolution}. For the IRDCs in our sample, we see no obvious correlation between infall velocities and the logarithm of the ratio of the {\NHn} column density to the CCS column density.

{\begin{figure}
    \includegraphics[width=.90\textwidth]{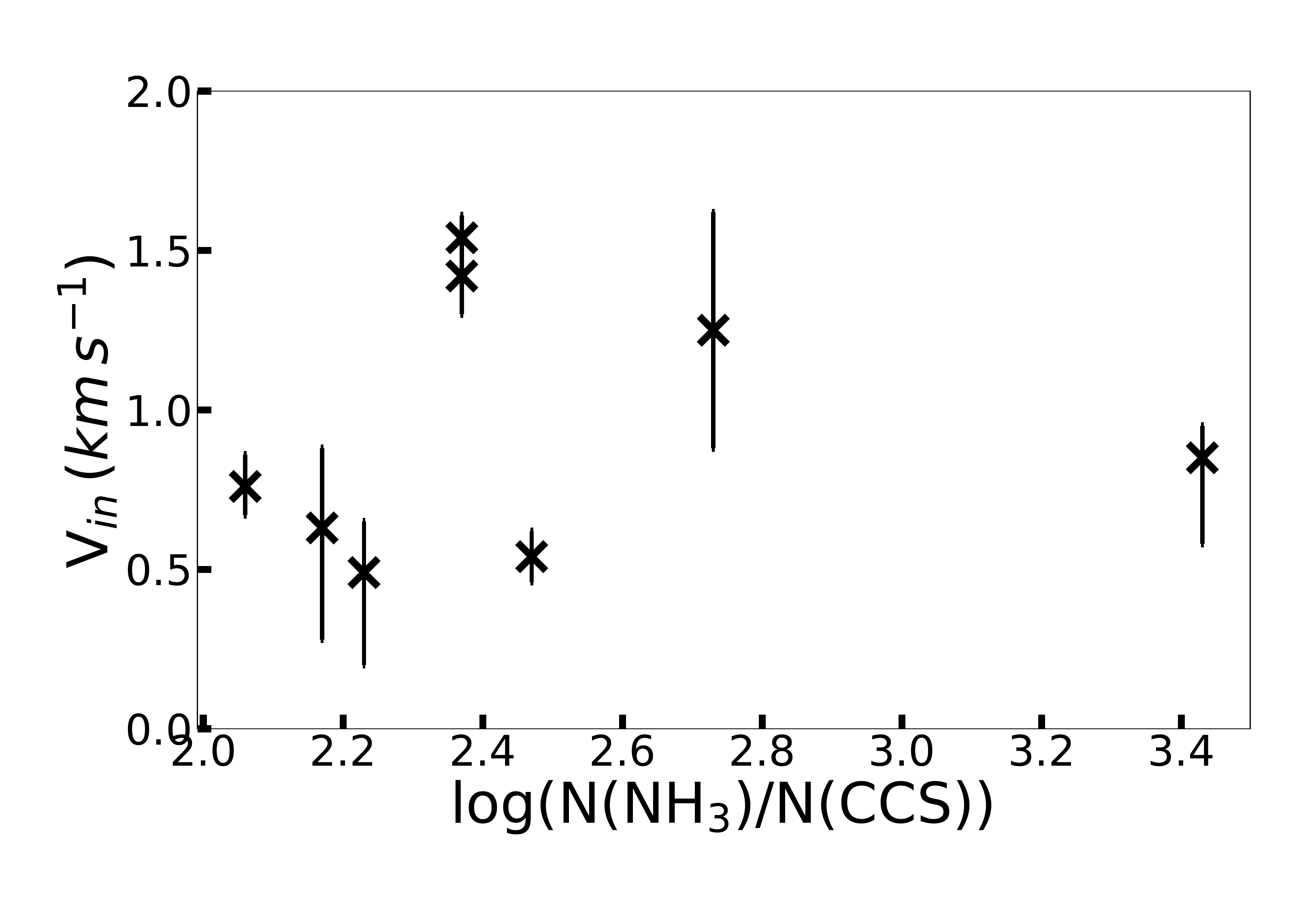}
    \hspace{1px}
    \caption{Infall velocities versus the logarithm of the ratio of the column density of {\NHn} to that of CCS. }
    \label{fig:evolution}
\end{figure}}

\section{Conclusions}\label{sect:concl}
We summarise our main conclusions as the following:

1. We investigate the infall properties in a sample of 11 IRDCs which present infall line profiles in a {\HCOn}$J\,$=1--0 survey by Peretto et al. (in prep). A total of 23 clumps were identified in {\HCOn}$J\,$=4--3 maps of theses IRDCs and they were observed in {\HCOn}$J\,$=3--2 using JCMT. The sizes and line luminosities of these IRDC clumps in {\HCOn}$J\,$=4--3 were calculated.

2. We compare the infall signatures traced by the three transitions of {\HCOn} towards the peaks of the 23 IRDC clumps and found that {\HCOn}$J\,$=3--2 can trace infall signature well in this {\HCOn}$J\,$=1--0 selected sample, while {\HCOn}$J\,$=4--3 shows the least blue asymmetric profiles in these IRDCs.

3. We used the HILL model to fit infall parameters based on the three {\HCOn} transitions profiles for the identified IRDC clumps. The infall velocities derived from {\HCOn}$J\,$=1--0 range from 0.5 to 2.7\,{\kms}, with a median of 1.0\,{\kms}, which are similar to the values derived from {\HCOn}$J\,$=3--2 and are more than 1.5 times larger than the values derived from  {\HCOn}$J\,$=4--3. 

4. The infall velocities and the mass accretion rates in the IRDC clumps in our survey are comparable to the values found in HMPO and UCHII samples. These IRDC clumps are more likely to form star clusters.

5. No prominent correlation between the infall velocities and the ratio of {\NHn} column density to CCS column density has been found in this sample.

\normalem
\begin{acknowledgements}
This work is supported by the National Natural Science Foundation of China (NSFC) grant No. 11988101, No. 11725313, No. 11721303, the International Partnership Program of Chinese Academy of Sciences grant No.114A11KYSB20160008, the National Key R\&D Program of China No. 2016YFA0400702. XIE J. J. acknowledges the support by the Chinese Scholarship Council (CSC) 
and the STFC China SKA Exchange Programme for support as a visiting PhD student in the United Kingdom.
G.A.F acknowledges support from the Collaborative Research Centre 956, funded by the Deutsche Forschungsgemeinschaft (DFG) project ID 184018867.
The authors wish to recognize and acknowledge the very significant cultural role and reverence that the summit of Maunakea has always had within the indigenous Hawaiian community.  We are most fortunate to have the opportunity to conduct observations from this mountain. XIE J. J. acknowledges the support and help from the JCMT observing team, especially Jim Hoge, Kevin Silva, Dan Bintley, and Maunakea Rangers. The {\sc Starlink} software \citep{starlink} is currently supported by the East Asian Observatory. This research made use of {\sc APLpy}, an open-source plotting package for Python\citep{aplpy2012,aplpy2019}.

\end{acknowledgements}
  
\bibliographystyle{raa}
\bibliography{bibtex}

\clearpage
\renewcommand{\thesection}{Appendix}

\subsection{ Observational Positions of RxA3m of {\HCOn}$J\,$=3--2 on the emission map of {\HCOn}$J\,$=1--0 with {\HCOn}$J\,$=4-3 contours. }\label{sect:level1}

\end{document}